\documentclass[12pt,draftclsnofoot,onecolumn]{IEEEtran}
\usepackage{graphicx,amsmath,amssymb}
\usepackage{subfigure}
\usepackage{citesort}
\usepackage{fancyhdr}
\usepackage{mdwmath}
\usepackage{mdwtab}
\usepackage{balance}
\usepackage{xcolor}
\usepackage{bm}
\usepackage{amsthm}
\usepackage{algorithm}
\usepackage{multirow}
\usepackage{flafter}
\usepackage{setspace}
\usepackage{cite}
\usepackage{algpseudocode}

\newtheorem{definition}{Definition}
\newtheorem{remark}{Remark}

\newtheorem{theorem}{Theorem}

\newcommand{\myb}[1]{\textcolor{black}{#1}}

\newtheorem{lemma}{Lemma}

\newtheorem{corollary}{Corollary}

\newtheorem{assumption}{Assumption}

\begin{document}
\title{A Reliable Reinforcement Learning for Resource Allocation in Uplink NOMA-URLLC Networks}
\author{Waleed~Ahsan,~\IEEEmembership{Graduate Student Member,~IEEE,}
        Wenqiang~Yi,~\IEEEmembership{Member,~IEEE,}
         Yuanwei~Liu,~\IEEEmembership{Senior Member,~IEEE,}
        and Arumugam~Nallanathan,~\IEEEmembership{Fellow,~IEEE}
\thanks{W. Ahsan, W. Yi, Y. Liu, and A. Nallanathan are with Queen Mary University of London, London, UK (email:\{w.ahsan, w.yi, yuanwei.liu, a.nallanathan\}@qmul.ac.uk).
\par Part of this work was submitted in IEEE Global Communications Conference (GLOBECOM), December, Spain, 2021 \cite{123456789}.}
}
\maketitle
\begin{abstract}
In this paper, we propose a deep state-action-reward-state-action (SARSA) $\lambda$ learning approach for optimising the uplink resource allocation in non-orthogonal multiple access (NOMA) aided ultra-reliable low-latency communication (URLLC). To reduce the mean decoding error probability in time-varying network environments, this work designs a reliable learning algorithm for providing a long-term resource allocation, where the reward feedback is based on the instantaneous network performance. With the aid of the proposed algorithm, this paper addresses three main challenges of the reliable resource sharing in NOMA-URLLC networks: 1) user clustering; 2) Instantaneous feedback system; and 3) Optimal resource allocation. All of these designs interact with the considered communication environment. Lastly, we compare the performance of the proposed algorithm with conventional Q-learning and SARSA Q-learning algorithms. The simulation outcomes show that: 1) Compared with the traditional Q learning algorithms, the proposed solution is able to converges within \myb{200} episodes for providing as low as $10^{-2}$ long-term mean error; 2) NOMA assisted URLLC outperforms traditional OMA systems in terms of decoding error probabilities; and 3) The proposed feedback system is efficient for the long-term learning process.
\end{abstract}


\section{Introduction}
As a key design feature for 5G and beyond 5G (B5G) networks, ultra-reliable low-latency communication (URLLC) attracts increased attention recently \cite{elbayoumi2020noma}. Thanks to the empirical advantages of URLLC that focuses on the transmission with finite blocklength (FBL) to support emerging applications, such as intelligent transportation systems\cite{xiang2020noma}, factory automation (industry 4.0), healthcare (remote surgery), virtual reality, and augmented reality. Such type of services requires guaranteed reliability, latency, and data rates for wireless communications \cite{zhang2020sparse} \cite{qian2020noma}. Note that these emerging real-time applications of the B5G network demands long-term optimizations for not only URLLC but also high data rates \cite{elbayoumi2020noma}. Different from orthogonal multiple access (OMA) schemes, NOMA is able to reduce the transmission delay for each user by providing additional access in the power domain, especially for extremely-high-frequency communications with low-rank channels~\cite{mu2020exploiting}. Therefore, NOMA offers the remedy for new rigorous requirements of the URLLC in two different ways \cite{gui20206g}. Firstly, NOMA enhances the spectral efficiency for URLLC. Secondly, due to simultaneously serving multiple users in the same resource block, NOMA is able to further reduce the latency for URLLC, especially for the scenario with massive users. Finally, to find an optimal balance between latency, reliability, and data rates, an intelligent-learning-based mechanism can be applied to dynamically operate the resource allocation in NOMA-URLLC networks. 

\subsection{Related Works}
\subsubsection{NOMA-URLLC}
Due to the simple design of OMA systems, most of the recent works concentrate on exploring URLLC with OMA instead of NOMA communications \cite{jiang2019packet,khoshnevisan20195g,xiang2019concurrent}. As URLLC uses FBL transmissions, the well-known Shannon capacity is no longer the accurate approximation of data rates \cite{she2017radio}. To characterise the FBL transmissions, the authors in \cite{polyanskiy2010channel} and \cite{yang2014quasi} designed a fundamental simulation framework. However, due to the design complexity of NOMA, evolving OMA-URLLC to NOMA-URLLC faces numerous challenges \cite{shahraki2021comprehensive}. In \cite{xiao2019downlink}, the authors designed downlink multiple-input multiple-output (MIMO)-NOMA framework for the URLLC networks. The design focused on a two-user case, which can be extended to multi-user scenarios to enhance the connectivity. Similarly, the work \cite{amjad2018performance} compared packet sizes, required signal to noise ratio (SNR), and delay constraints for the NOMA-URLLC systems for the two-user case. The simulation outcomes showed that the required SNR is inversely proportional to the quality of service exponent and error probabilities. The authors in \cite{kotaba2019improving} designed a static multi-user NOMA-URLLC framework based on hybrid automatic repeat request re-transmissions. In this framework, fixed user connectivity was simulated for different network settings to show the performance of NOMA-URLLC. The results section showed that the performance of NOMA-URLLC is better than OMA-URLLC schemes in all settings. The authors in\cite{chen2019optimal} provided a detailed numerical framework for NOMA-URLLC using a dynamic programming approach. Based on error probabilities and the maximum transmit power, the simulation outcomes proved that NOMA-URLLC systems perform efficiently in all discussed cases. Finally, in \cite{8933345}, the authors presented the advantages of NOMA based URLLC transmission use cases over OMA-URLLC systems. Notably, existing works do not address data rates, reliability, and latency constraints properly for a long-term basis.
\subsubsection{AI-NOMA}
In comparison to the model-free methods, the traditional model-based approaches are not suitable for simultaneously controlling multiple network parameters. Due to frequent fluctuations in wireless systems, a significant amount of model-free frameworks are being investigated \cite{maraqa2019survey}. For example, the authors in \cite{huang2020deep} proposed communication deep neural network (CDNN) to optimise the sum utility of the MIMO-NOMA system. The simulation results showed that CDNN outperforms the conventional methods by providing efficient system utility. In this model, the stochastic gradient descent (SGD) optimization was used to minimise the training loss with fully connected layer of linearly rectified (ReLU) activations. Similarly, inspired by the grant-free NOMA, the authors in \cite{ye2019deep} designed a DNN framework to address low latency access for tactile internet of things (IoT) system. The performance outcomes showed that the proposed model provides better results than multi-user shared access (MUSA) technique. \myb{A meta deep learning based framework is also proposed in \cite{9310298} to handle dynamic network parameters for the supervised learning scheme. In which a general DNN model is trained on pre-generated data-set to learn the communication network settings. In simulation outcomes meta learning based model free system was able to learn the system efficiently than traditional learning technique.} In a recently proposed framework \cite{ye2020deepnoma}, the authors also used DNN to design the deepNOMA system. According to their simulation outcomes, in most of the cases the deepNOMA framework performs better than other schemes like conventional OMA and model-based NOMA systems. Different from others, the authors in \cite{fu2019dynamic} exploited DNN for the NOMA based caching scheme for the dynamic power control. The final results were verified using Monte Carlo simulations that showed the learning-based algorithm performs better than traditional model-based schemes. Interestingly, all the existing DNN based models require huge data-sets for training that is difficult to be acquired for practical NOMA systems. Therefore, apart from other advantages, the similar models suffer from the same limit, where the training is only based on data-sets instead of the actual environment. This limit may result in an unreliable decision since the real-time feedback is not effectively considered. Hence, practical NOMA systems, especially for URLLC, require a reliable model-free design that can learn from actual wireless network environments.         

\subsection{Motivations}
\myb{For emerging NOMA-URLLC applications, long-term reliable optimization of resource allocation to improve the latency, reliability, and data rates is challenging but important \cite{parvez2018survey,azari2019risk,she2020tutorial}. Because in long-term optimization process the learning strategy is developed to predict future resource allocations. Therefore, the long-term strategy is based on the real experience of the agent (BS) that is gained by interactions with the physical environment to ensure the stability and reliability. As discussed in the related works, the existing solutions for URLLC are mainly based on conventional OMA or model-based NOMA (short-term) conventional schemes. A new solution with prompt real-time responses is urgently needed to enhance the reliability in learning based resource allocation process. Recently, human-level control accuracy over the continuous control is achieved in the game of GO by the use of reliable learning algorithms, such as $(n)$-step reinforcement learning (RL) algorithms and google deepmind \cite{hernandez2019understanding,silver2016mastering,mnih2015human}. Therefore, such type of techniques are suitable to handle dynamic networks. Compared with other RL techniques, e.g., traditional Q-learning \cite{watkins1992q}, multi-arm bandits \cite{sutton2018reinforcement},  and DNN's, the $(n)$-step RL is a promising direction to learn long-term reliable policies. Since, $(n)$-step RL performs the future actions/decisions depending on the knowledge from multiple ($(n)$-step) interactions for the purpose of enhanced efficiency. In other words, as the process of RL algorithms is based on the rewards and punishments for this reason with the help of ($(n)$-step) optimizations the agent finds the long-term effective allocation mechanism to avoid the punishments. Consequently, the learned policy is suitable for long-term in dynamically changing wireless network scenarios. So that most recent models in wireless networks are based on multi-step learning \cite{ahsan2021resource,shao2020significant}.} 
\par \myb{Motivated from the aforementioned discussions, for time-varying network environment, we consider to design a model-free ($(n)$-step) learning framework to address long-term resource allocation problem for URLLC with NOMA transmissions. Moreover, a multi-user NOMA scheme is considered to further enhance the performance.}       
\subsection{Contributions and Organization}
In the result of the above motivations, we design a reliable framework to learn NOMA-URLLC uplink communication environments in terms of the long-term performance. With a reliable $(n)$-step learning framework, we first formulate the considered problem as the reward maximization issue by jointly optimising the resource allocation and error probabilities. Then, we propose the Deep SARSA$-\lambda$ algorithm for the long-term optimization with rewards. Lastly, to enhance the reliability, we utilise the sparse neuron activation that are based on the ReLU and backward trace table. The detailed summary of our contributions are listed as follows:     
 \begin{itemize}
\item We design a 2D model-free clustering framework to serve NOMA-URLLC users dynamically based on available network resources. Considering this framework, we devise a mean error minimization problem to handle URLLC constraints. These constraints consider long-term variables in the proposed NOMA-URLLC systems, such as the packet size of users, the number of users and channel gains. To portray fully online behaviour,  these parameters are alterable at each time step.
\item We propose a reliable deep RL (DRL) algorithm, as deep SARSA$-\lambda$ with dynamic $\lambda$ and reliability tracing, to address the long-term reliability problem. The deep neural network is used for forward tracing to increase short-term reliability. Long-term reliability is handled by backward tracing via dynamic $\lambda$.    
\item We design a reward function $D_r(t)$, that is used to train the BS for optimal resource allocation under URLLC constraints. Additionally, to avoid simple (non-practical) case with a static reward function that is predefined by the designer. The proposed reward function $D_r(t)$ is based on the instantaneous system output (system utility) at each timestep $(t_s)$, rather than conventional user defined values. Furthermore, to increase the exploration safety and reliability we design backwards trace table that assists the DRL agent to avoid unreliable state visits.  
\item We demonstrate that: 1) Based on \textcolor{black}{time varying} properties, reliable resource allocation can be performed simultaneously in time varying systems; 2) According to the proposed framework, the reward function provides optimal trade-off for the long-term error (Fig. \ref{2}) and the sum rate (Fig. \ref{1}) by best convergence; 3) For deep SARSA$-\lambda$, the long-term mean error probability is proportional to the network density and packet size; 4) For DNN prediction accuracy, the mean squared error (MSE) is also proportional to the network density and packet size; and 5) NOMA is more suitable for URLLC due to providing the efficient spectral utility with low decoding error probabilities as compared to OMA systems.   
  \end{itemize}
\begin{table}[htb!]
\small
\centering
\caption{Table of notations}
\label{tab11}
\begin{tabular}{|l||l|}
\hline 
   \textbf{Symbol} &  \textbf{Definition}  \\ \hline
   \text{$N_s,s_j$} &  Number of sub-channels (NOMA clusters), symbol of sub-channels (NOMA clusters)  \\ \hline
   \text{$N_u,u_k$} &  Number of users, symbol of users  \\ \hline	
   \text{$\Upsilon_s$}  &Set of sub-channels (NOMA clusters) \\ \hline
   \text{$\Upsilon_u^{j}$,$u^{j}_k$} & Set of users connected to BS via sub-channel $s_j$, user $k$ in the set $\Upsilon_u^{j}$   \\ \hline
      \text{$\Upsilon_m,m_k$}  &Set of block sizes with Block size $m$ for the user $k$ \\ \hline
   \text{$c^{j}_{k}(t)$} & Clustering variable for user $u_k$ connecting to BS via sub-channel $s_j$ at time $t$  \\ \hline
   \text{$p^{j}_{k}(t)$} & Transmit power for user $u^{j}_k$ at time $t$  \\ \hline
   \text{$g^{j}_{k}(t)$} & Channel gain for user $u^{j}_k$ at time $t$\\ \hline
   \text{\myb{$\sigma^2(t)$}} & Additive white Gaussian noise at time $t$ \\ \hline
   \text{$D$} & Total number of the transmission bits for each user \\ \hline
   \text{$M$} & Maximum block size \\ \hline
   \text{$V$} & channel dispersion \\ \hline 
   \text{$\Psi(\gamma^{j}_{k}(t) ,M,D)$} & Q value Input calculation function  \\ \hline
   \text{$\varepsilon^{j}_{k}(t)$} &  Decoding error probability \\ \hline
   \text{$\varphi$} & Set of error probabilities \\ \hline
   \text{$\gamma^{j}_{k}(t)$} & Instantaneous SINR for user $u^{j}_k$ at time $t$\\ \hline
   \text{$R^{j}_{k}(t)$} & Instantaneous data rate for user $u^{j}_k$ at time $t$\\ \hline
   \text{$R^{th}_{k}$} & Rate requirement for the SIC process of user $u^{j}_k$ \\ \hline
   \text{$D_r(t)$} & reward function \\ \hline
   \text{$T$} & Duration of the considered long-term communication \\ \hline
   \text{\myb{$\tau(s,a)$}} & Reliability tracing table \\ \hline
   \text{\myb{$Q(s,a)$}} & Q table \\ \hline
   \text{\myb{$\varrho(\vartheta)$}} & activation function with input values $(\vartheta)$\\ \hline
   \text{\myb{$\mathbb{S}$}} & State space \\ \hline
   \text{\myb{$\mathbb{A}$}} & Action space \\ \hline
   \text{\myb{$\mathbb{P}$}} & Transition probabilities \\ \hline
   \text{\myb{$\pi$}} & q learning policy \\ \hline
    \text{\myb{$\kappa$}} & Experience memory \\ \hline
   \text{\myb{$Q(.)$}} & Error function \\ \hline
   \text{\myb{$\lambda$} }& Learning step  \\ \hline
   \text{$\mathbf{C}_t$, $\mathbf{P}_t$} & Matrix for clustering parameters, matrix for transmit power \\ \hline
   \text{$\mathbf{ \theta}$} & DQN error/loss \\ \hline
   \text{\myb{$P_{s(t) \rightarrow s(t+1)}$}} & \myb{state transition probability} \\ \hline
\end{tabular}
\end{table}
\begin{figure}[htb]
  \centering
  \includegraphics[scale=.55,keepaspectratio]{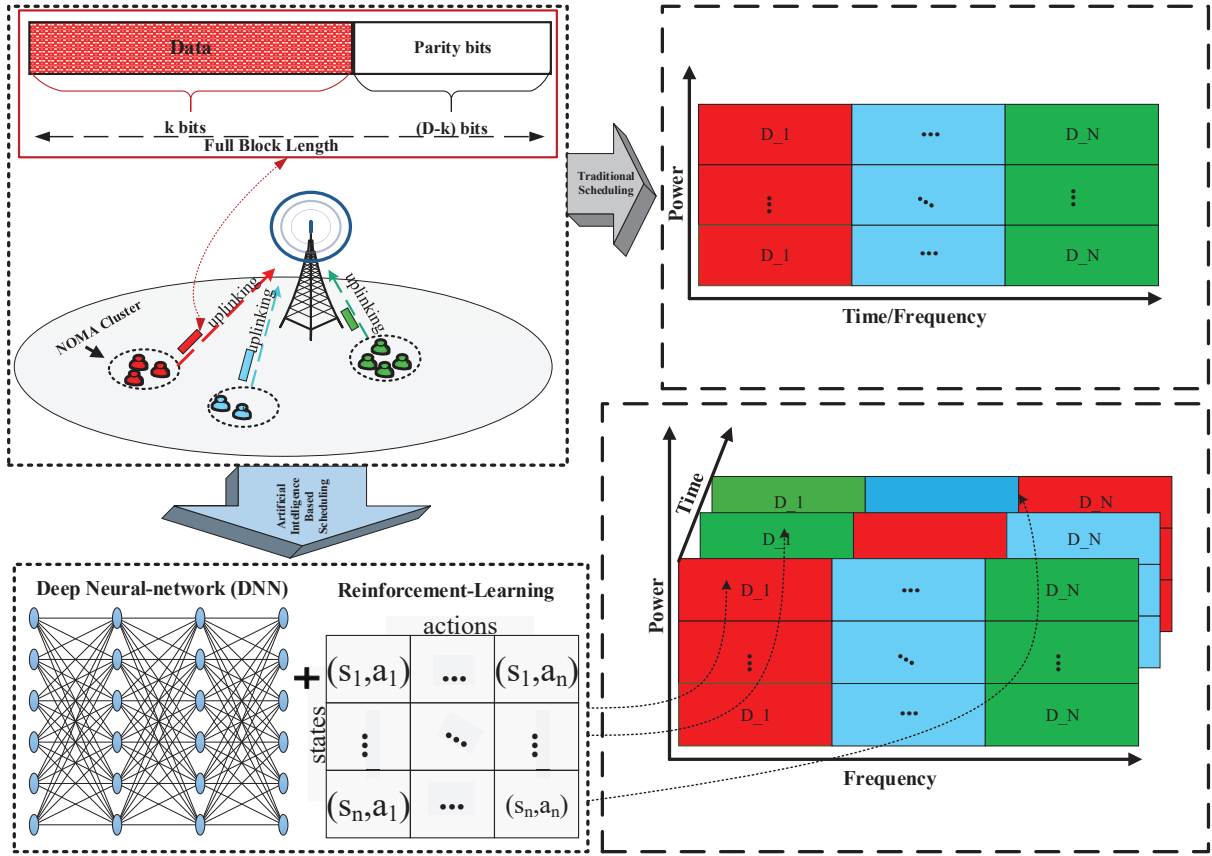}\\
  \caption{Illustrating NOMA-URLLC uplink resource allocation as a model-free optimization problem. The top-left sub-figure shows details for the wireless network (uplink NOMA-URLLC) including NOMA clusters highlighted in (red, green and blue colours), a base-station, NOMA users, and the URLLC packet structure. The top-right sub-figure presents fixed resource allocations under traditional strategies. The bottom sub-figures shows the resource allocation strategies (on the right) using the deep reinforcement learning methods (on the left).}\label{intro}
\end{figure}																						
\section{Network Model and Problem Formulation}
\subsection{Network Settings}
This paper investigates NOMA-aided uplink URLLC with short data blocks as shown in Fig. 1. One BS is located at the center of the considered cell. The BS communicates with $N_u$ users via $N_s$ orthogonal sub-channels. Both the BS and users are assumed to be equipped with a single antenna. The set of data block sizes $m_k\in\Upsilon_m$ $(k\in[1,N_u])$ for all users $\Upsilon_m=\lbrace{m_1,m_2,\cdots,m_{N_u}\rbrace}$ are assumed to be the same, which consists of $D$ bits. The transmission of these blocks is subject to a latency constraint, i.e., for each sub-channel the transmission of one data block should be finished within $M$ seconds per unit bandwidth. The sub-channels are indexed by $\Upsilon_s = \{s_1,...,s_{N_s}\}$. Regarding users, the set for users severed by the BS through a sub-channel (i.e., a NOMA cluster) $s_j\in\Upsilon_s$ $(j\in[1,N_s])$ is defined as $\Upsilon_u^{j}=\{u_1,...,u_{N_{u}^{j}}\}$, where $N_u^{j}$ is the number of the users connected to the BS via the sub-channel $s_j$ and $\sum\limits_{j = 1}^{{N_s}}N_u^{j}=N_u$. The defined notations in this system model are listed in the TABLE \ref{tab11}.
\subsection{Signal Model}
 In NOMA, two or more users are able to access the same resource block (time/frequency). Therefore, each sub-channel has $N_u^{j} \ge 2$ \cite{kiani2018edge}. To simplify the analysis, we assume the BS contains perfect CSI of all users. Based on such CSI, BS is capable to \textcolor{black}{simultaneously} optimise the sub-channel allocation for active users in long-term communications. For an arbitrary user $u_k$, we define its active index variable at time $t$ as $c^{j}_{k}(t)=1$ when the user $u_k$ is served by the NOMA cluster $s_j$. Similarly, $c^{j}_{k}(t)=0$ means the users are silent in the sub-channel $s_j$. The set of clustering parameters is defined as $\mathbf{C}_t$ and $c^{j}_{k}(t) \in \mathbf{C}_t, \forall j,k$. The size of $\mathbf{C}_t =  N_u \times N_s$ in this model.

In a NOMA cluster $s_j$, when the BS first receives the transmitted messages from the connecting users from $\Upsilon_u^{j}$, the signal of each user is decoded by applying SIC~\cite{mu2020exploiting}. 
Additionally, we assume that the decoding order in this model is the reverse of the channel gain order~\cite{8680645}. In a time slot $t$, the instantaneous signal-to-interference-plus-noise ratio (SINR) for the intra-cluster user $u^{j}_k\in\Upsilon_u^{j}$ is given by 
\begin{align}\label{a}
\gamma^{j}_{k}(t)= \frac{c^{j}_{k}(t)p^{j}_{k}(t) g^{j}_{k}(t)}{\sum\limits_{{k'=1}}^{k-1}{c^{j}_{k'}(t)p^{j}_{k'}(t) g^{j}_{k'}(t)}+\sigma^2(t)},
 \end{align}

\noindent where the SINR for the last user is defined as
\begin{align}\label{a2}
\gamma^{j}_{1}(t)= \frac{c^{j}_{1}(t)p^{j}_{1}(t) g^{j}_{1}(t)}{\sigma^2(t)},
 \end{align}
 where $p^{j}_k(t)$ is the transmit power of the user $u_k$ and the set of transmit power is given by $\mathbf{P}_t$ $(p^{j}_{k}(t) \in \mathbf{P}_t, \forall j,k)$  and the size of $\mathbf{P_t}$ obeys that \myb{$\mathbf{|P_t|}=N_u$}. The $\sigma^2(t)$ is the additive white Gaussian noise (AWGN) noise factor.
\subsection{Achievable Data-rate with Finite Block Length}
In uplink finite block-length NOMA transmission, the decoding of user $u_k$ is based on the SIC process of its previous user $u_{k+1}$. For all of the users in the finite block-length transmissions the decoding error probabilities are denoted by the set $\varphi=\lbrace{\varepsilon^{j}_{1}(t),\varepsilon^{j}_{2}(t),...,\varepsilon^{j}_{{N_{u}^{j}}}(t)}\rbrace$. Additionally, if the data rate of successfully completing the SIC process is $R_{k}^{th}$, according to \cite{8933345}, the decoding rate (in bps/Hz) of user $u^{j}_{k}$ obeys 
\begin{align}\label{ur1}
 R^{j}_{k}(t)\approx \log_2 (1+\gamma^{j}_{k}(t) )-\sqrt{V/M}\frac{{{Q}^{-1}}(\varepsilon^{j}_{k}(t) )}{\ln 2},
\end{align}
where
\begin{align}\label{ua}
 V=1-{{(1+\gamma^{j}_{k}(t) )}^{-2}},
 \end{align}
\myb{If $R^{j}_{k}(t) < R_{k}^{th}$ the target rate threshold $R_{k}^{th}$, the decoding of all the users $u_{k}, ... ,u_1$ fails, namely $R^{j}_{k}(t)=...=R^{j}_{1}(t)\equiv 0$. ${Q}^{-1}(.)$ is the inverse function of the following equation:} 
\begin{align}\label{ub}
Q(\zeta)=\frac{1}{\sqrt{2\pi }}\int_{\zeta}^{\infty }{{{e}^{\frac{{{t}^{2}}}{2}}}}dt,
\end{align}
Based on the results in \cite{yang2014quasi}, the expression for the decoding error probability and the channel dispersion with the Rayleigh fading can be expressed as
\begin{align}\label{uc}
\varepsilon^{j}_{k}(t) =Q(\Psi(\gamma^{j}_{k}(t) ,M,D)),
\end{align}
where
\begin{align}\label{ud}
\Psi(\gamma^{j}_{k}(t) ,M,D)=\ln 2\sqrt{\frac{M}{V}}({{\log }_{2}}(1+\gamma^{j}_{k}(t) )-\frac{D}{M}),
\end{align}
\noindent and $(\frac{D}{M})$ is the required decoding rate for satisfying the latency constraint.
\subsection{ Problem Formulations}
\myb{In this section, we construct uplink-NOMA-URLLC based optimization problem. This work minimizes the decoding errors, since reducing decoding errors is capable for both enhancing the reliability and decreasing the latency. The function to minimize the average decoding errors for long-term communications with a period $T$ is as follows:}
\myb{
\begin{subequations}\label{Rx}
\begin{align}\label{h}
\underset{\mathbf{C_t,P_t} }{\min} \ \ \ & \mathbb{E}\left[ \sum_{t_s=1}^{T}\sum_{j=1}^{N_s}\sum_{k=1}^{N_{u}^{j}}\,{\varepsilon_k^j} (t_s)\right] \\
\mathrm{s.t:}\ \ \ \label {i}&p^{j}_1g^{j}_1(t_s) \leq, ... ,\leq p^{j}_{{N_{u}^j}}g^{j}_{{N_{u}^j}}(t_s),\ \forall j, t_s,\\
&\label{j}{\myb{\sum_{k=1}^{N_u^{j}}{c^{j}_{k}(t_s) p^{j}_{k}(t_s) \le P_s },\forall j, t_s}},\\
&\label{k}R^{j}_{k}(t_s) > \myb{R_{k}^{th}},\ \forall k, t_s,\\
&\label{l} {2}\le\sum_{k=1}^{N_{u}^{j}}{c^{j}_{k}(t_s)\le N_u,} \ \forall j,t_s, \\
&\label{o1}{\textcolor{black}{\sum_{j=1}^{N_s}c^{j}(t_s) \in \{1,0 \}}}, \ \textcolor{black}{\forall j, t_s,}
\end{align}
\end{subequations}}
\myb{where \eqref{i} shows the channel ordering for the network user based on the perfect CSI. \eqref{j} is to impose the power constraint of each sub-channel. \eqref{k} ensures all clustered NOMA users can be successfully decoded for maximizing the connectivity by ensuring that the rate $R^{j}_{k}(t_s)$ of each user is more than the target data-rate $R_{k}^{th}=0$. \eqref{l} limits the number of clustered users for the entire system and each sub-channel. The constraint \eqref{o1} indicates that each user belongs to only one cluster. All of these constraints in our coding are checked in the form of sequential steps, if the value is out of range from (8b-8e) then that try is discarded to check it again. The joint optimization problem \eqref{h} is a non-convex and NP-hard problem. Therefore, all the sub-problems (i.e., resource allocation and \textcolor{black}{simultaneous} user connectivity) in \eqref{h} are also non-convex and NP-hard, especially when considering \eqref{j} and \eqref{l}. The detailed proof under traditional NOMA has been provided in \cite{8807386}. However, by jointly considering the resource allocation and the finite packet conditions, the problem in \eqref{h} becomes more complex. As a solution, the modern artificial intelligence techniques are known to provide long-term reliable solutions for such type of problems. Therefore, we propose the following model to handle this issue.}
\section{Reliable Resource Allocation for URLLC-NOMA}
In this section, we propose an intelligent Deep SARSA$-\lambda$ based reliable resource allocation algorithm to assist the BS for minimizing the error probability for uplink NOMA-URLLC. Now, we present the basic concepts and problem formulation for the deep SARSA-$\lambda$ followed by reliable estimations, algorithmic details, and analysing the proposed algorithm.   
\subsection{The Reliable Learning Policy with SARSA-$\lambda$}
\subsubsection{Multi-step introduced reliability}
\myb{Unlike the traditional one step RL methods such as, Q-learning and SARSA Q-learning. The deep SARSA$-\lambda$ algorithm performs $(n)$-step optimization process for dynamically changing networks (such as NOMA-URLLC with simultaneous user connectivity). This property ensures that deep SARSA$-\lambda$ is more reliable than conventional policy learners, which is discussed in \textbf{Remark 1}. Before that, we first introduce the Markov Reward process.}
\begin{definition}
(Markov Reward Process:) A Markov Reward Process (MRP) is based on the tuple $\mathbb{\lbrace{S,A,R,P\rbrace}}$. For every time slot $t$, the element $\mathbb{\lbrace{P\rbrace}}$ is a probability distribution for the movement of an agent from state $s(t) \in \mathbb{S}$ to $s(t+1)$ that is conditioned on the action $a(t)$ to maximise the associated reward $D_r(t)$, that is expressed as $D_r(t)\mathbb{:S \times A \times P \Rightarrow R}$.
\end{definition}
\begin{figure}[htb]
  \centering
  \includegraphics[scale=.5,keepaspectratio]{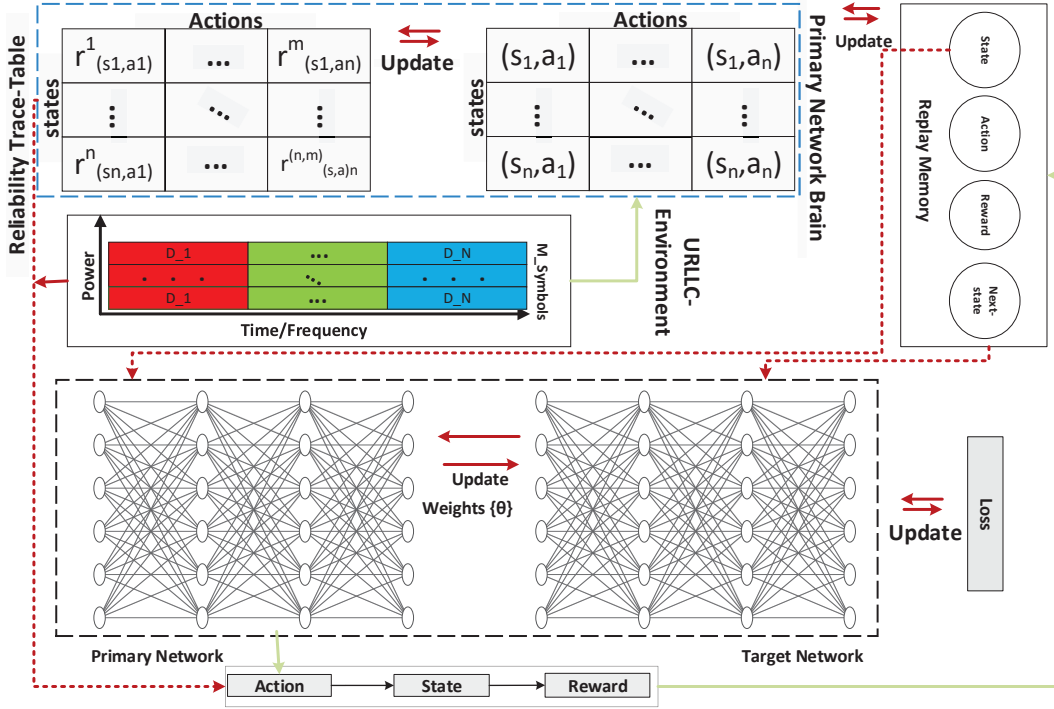}\\
  \caption{Presenting the structure of the proposed reliable resource allocation algorithm. Illustrating the communication environment where the Deep SARSA$-\lambda$ is invoked. Additionally, it shows the reliability assistance mechanism between primary DNN and the target DNN to maximise the reward by utilizing replay memory to minimise the DNN loss function.}\label{frame}
\end{figure}
\begin{remark}\label{r11}
\myb{For the reliable policy $\pi^*\mathbb{:S \Rightarrow A}$, to balance the learning process for huge $\mathbb{S}$ in the time varying networks. The SARSA-$\lambda$ algorithm adopts deep neural network (DNN) and the dynamic $\lambda$ to keep track of $(t-n)$ allocations. Due to this $(n)$-step tracing capability, the SARSA-$\lambda$ achieves the efficient path and enhanced reward performance by choosing the most efficient set of actions in the future. }  
\end{remark}
\noindent \myb{Therefore, based on the one-step learning mechanism, the traditional agent/s cannot guarantee the efficient policy by depending only on $a(t),s(t),r(t)$ and the current allocation strategy. The detailed reason is offered in the following remark.} 
\begin{remark}
\myb{For the time varying network environments, to enhance the network reliability, the $(n)$-step MRP's SARSA-$\lambda$ eliminates the less reliable allocations updating trace table with '$0$', as shown in \eqref{tr}. However, for the long-term \textcolor{black}{time varying} communication environments, conventional one-step expectation return MRP's Q-learning and SARSA Q-learning perform allocations without any prior knowledge. Therefore, the final allocation strategy for conventional approaches is not efficient.}  
\end{remark}
\myb{\subsubsection{Dynamic $\lambda$ introduced reliability evaluations}
The dynamic $\lambda$ determines the reward performance of each step of an agent leading towards the effective policy, that is enabled by trace table $\mathbf{\tau}(s,a)$. Different from one-step expectation return MRP (Q-learning and SARSA Q-learning), the $(n)$-step MRP (SARSA-$\lambda$) benefits from combined properties of the dynamic $\lambda$ and the tracing table $\mathbf{\tau}(s,a)$. Because the tracing table $\mathbf{\tau}(s,a)$ is updated using dynamic $\lambda$ over $(n)$-steps, that is defined as $\lambda \in [0,1]$. When the value of $\lambda$ is large, it indicates the importance/reliability of current state and action pairs is high, and vice versa. Similarly, due to the dynamicity of $\lambda$ and tracing property, the agent/s becomes $(n)$-step learner/s. Therefore, the magnitude of parameter $\lambda$ assists the agent/s (BS) at time slot $(t)$ to prioritise its short-term experience. Hence, it ensures that the agent (BS) is more reliable by careful assistance. Furthermore, if the $\lambda=0$ then the agent decides action $a(t)$ without any experience that results in less reliability for the future actions.
\begin{remark}
For the fixed $\lambda$, the agent performs MRP based on the current Q-values and current reward only. However, when the $\lambda$ is dynamic the agent performs short-term experience tracing based on multiple time-slots to allocate resources in the future. Hence, by keeping the $\lambda$ dynamic the agent knows the summary of the past $(n)$-steps for the long-term future allocation strategies.   
\end{remark}}
\subsubsection{DNN introduced long-term control}
\myb{As it is known that general MRP consists of observations/states, actions, rewards, policies, environment,  and RL agent/s. The agent consists of Q-table and DNN $\theta$ weights that represent the brain of one-step learning agents. Similarly, the probabilities for each state and action pair are learned via Q-table $\mathbf{Q}(s,a)$ and DNN $\theta$ weights. Therefore, Q-table $\mathbf{Q}(s,a)$ and DNN $\theta$ weights are updated during the MRP. The overall MRP is achieved in two main parts namely, short term and long-term learning.} 
\par \myb{In the short-term learning process, the agent performs actions to maximise the immediate reward feedback. That helps the agent/s to learn the long-term policy in the form of a set of actions for each state of an agent, represented with the help of probabilities stored in Q-table $\mathbf{Q}(s,a)$ and DNN $\theta$ weights. Consequently, in the result of the long-term optimal policy, the RL agent is expected to gain more long-term rewards. This process is only for the traditional one-step design where the agent can only learn and update probabilities once for each timestep $(t_s)$. Due to this, the  one step learning agent can revisit the useless states during the MRP that costs poor convergence. On the other hand, different from one step learning the $(n)$-step design combines the brain of the agent with trace table $\mathbf{\tau}(s,a)$ to learn multiple actions and update multiple the probabilities at once. Hence, on the cost of the trace table $\mathbf{\tau}(s,a)$ the $(n)$-step learning agent increases the speed of convergence and the reliability for the long-term learning policy.}
\begin{remark}
To speed-up learning the reliable allocation policy $\pi^*$, in addition to $(\theta)$ weights and tracing table $\mathbf{\tau}(s,a)$ the Deep SARSA-$\lambda$ agent (BS) is equipped with the short-term replay memory. The replay memory consists of online learning traces from the real-time network environment and tracing table $\mathbf{\tau}(s,a)$. Consequently, an experienced Deep SARSA-$\lambda$ agent performs efficient actions $a(t)$. As a result, the future action $a(t+1)$ selection policy $\pi^*$ becomes more reliable for an agent to converge faster by avoiding exploration visits to less reliable states $s$.  
\end{remark} 

Based on the above discussions, the following sub-section provides the design details for proposed resource allocation algorithm. The design consists of the basic design of RL agent, wireless network environment, state space $\mathbb {S}$, action space $\mathbb {A}$, reward function $D_r(t)$, transition probability $T(s (t),a (t),s (t+1),a (t+1))$, DNN design, and tracing mechanism.  
\subsection{Design of the SARSA-$\lambda$}
\begin{itemize}
\item \underline{State Space $\mathbb {S}$:}
\myb{is a state space consists of finite set having dimensions $N_s$ containing $ N_u^{N_s}$ total number of states. Each state represents one sub-set of 2D associations among users and sub-channels for the BS.}
\item \underline{Action Space $\mathbb {A}$:}
\myb{ is an action space consists of a finite set of actions to move the agent in a specific environment, the size of $\mathbb {A}=(N_{s}-1) \times N_s$. Actions in this model are $[-1,0,+1]$. The `$-1$' is to reduce one user from the current state of the agent (BS). Similarly, `$+1$' shows an increment in one user in the current state of the agent (BS). The last action `$0$' represents no change in the current state of the agent (BS). For example for one swap operation for both sub-channels and power levels, if the current state of an agent is indicating that $N'$ users are connected via sub-channel $j$ with $\rho_{N'}$ power levels used from it's pool $\rho_{j}$ and the sub-channel $j+1$ consists of $M'$ users with $\rho_{M'}$ power levels used from it's pool $\rho_{j+1}$. Note that the size of each pool is same and denoted by $\rho_{j+N_s}=N_u$. Similarly, when the agent performs action `$1$' for the sub-channel $j$ and `$-1$' for the sub-channel $j+1$ to swap one user from sub-channel $j+1$ to sub-channel $j$ and also swap and update the power pools from $\rho_{M'-1}$ to $\rho_{N'+1}$. The result of this action is that in the sub-channel $j$, the number of users is from $N'$ to $N'+1$ with new power level utility $\rho_{N'+1}$, while in the sub-channel $J+1$, the number of users is from $M'$ to $M'-1$ with released power level $\rho_{M'-1}$.  The last action `$0$' means no swap operation. Therefore, in results of the last action `$0$' the current state of the agent does not change (add/subtract). In this way the agent finds the final combination of state and action pair. } 
\item \underline{reward Function $D_r(t)$:}
In the MRP, the agent tries to maximize the sum reward for each time slot $(t)$ for the transition from current state action pair $\{s(t),a(t)\}$ to the future state and action pair $\{s(t+1),a(t+1)\}$. We use instantaneous sum rate as a reward for an agent during each time slot $(t)$. The reward function is as follows:
\begin{align}\label{dr}
      \myb{D_r(t)=
       \begin{cases}
       \sum\limits_{k=1}^{N_{u}^{j}}R_{k}^{j}(t),& \text{$if \ \mathbb{E}\left[\sum\limits_{j=1}^{N_s}\sum\limits_{k=1}^{N_{u}^{j}}\,{\varepsilon_{k}^j} (t-1)\right]\ge \mathbb{E}\left[\sum\limits_{j=1}^{N_s}\sum\limits_{k=1}^{N_{u}^{j}}\,{\varepsilon_{k}^j} (t)\right]$} \\& \text{ $and$ \ $\sum\limits_{j=1}^{N_s}\sum\limits_{k=1}^{N_{u}^{j}}{c_k^j(t-1)=\sum\limits_{j=1}^{N_s}\sum\limits_{k=1}^{N_{u}^{j}}{c_k^j(t)}}$}\\
       0,& $otherwise.$
		\end{cases}}
\end{align}
\myb{where the reward $D_r(t)$ is equal to instantaneous sum rates $\sum\limits_{k=1}^{N_{u}^{j}}R_{k}^{j}(t)$ if the expectation sum of the function $\varepsilon_{k}^j (t)$ for the time slot $(t)$ is less than the previous time slot $(t-1)$ and $0$ for rest of the cases. While keeping all the connected with $\sum\limits_{j=1}^{N_s}\sum\limits_{k=1}^{N_{u}^{j}}{c_k^j(t-1)=\sum\limits_{j=1}^{N_s}\sum\limits_{k=1}^{N_{u}^{j}}{c_k^j(t)}}$ conditions.}
\item \underline{Value Function $V_\pi$:}
In MRP the value function ${{V}_{\pi}}$ shows the expectation return in result of each state and action pair.
\begin{align}
{{V}_{\pi}}=\mathbb{E}\{{D_r(t)}|{s(t)}=s,a(t)=a\},
\end{align}
where ${{V}_{\pi}}$ denotes the value function that is similar to long-term expected return.
\item \underline{Transition Function $T(s (t),a (t),s (t+1),a (t+1))$:}
The state and action transition function shows the expectation probabilities, where the agent changes its current state to the future for the search of optimal state and action pair. As we can see from the Fig. \ref{frame}, the replay memory is updated after each transition of current state of an agent to the future state in result of the action $a$. The transition  function for the proposed method is as follows.
\begin{align}
\myb{p(s',D_R|s,a)=\Pr \{{s(t)}={s'},{D_r(t)}=D_R(t)|{s(t-1)}=s,{a(t-1)}=a\}},
\end{align}
\myb{where $\Pr$ is the probability distribution for the transition of state $s$ and action $a$ to the next state  $s'$ of the environment. The $(|)$ is to show the conditional probability.}
\item \underline{Q-Table $\mathbf{Q}(s,a)$:}
The Q-Table is a primary brain of an agent (BS), that keeps track of all the Q-values for each state and action pair, denoted by $\mathbf{Q}(s,a)$. The rows of a table means total number of states that is 2D associations in our case and the total number of actions are presented with the number of columns, as different swap operations. 
\item \underline{State and Action Trace-Table $\mathbf{\tau}(s,a)$:}
\myb{The discount value $\gamma$ and $\lambda$ are mainly used to enhance the packet transmission reliability with the help of Trace-Table $\mathbf{\tau}(s,a)$, due to multi-step return capability. The trace memory helps to improve the learning of an agent/s at each time slot $(t)$, the trace memory is denoted with $\mathbf{\tau}(s,a)$ that is initialised with 0 for every episode. Similarly, in this way an agent assisted with the trace memory $\mathbf{\tau}(s,a)$ enhances the long-term reliability of the URLLC system. To reflect fully online behaviour, in this model we use replacing reliability traces and the reliability trace value update, which can be expressed as follows:
\begin{align}\label{tr}
\myb{\mathbf{\tau}(s,a)=\begin{cases}
       \gamma\lambda\tau(s(t-1),a(t-1))+1, &   \text{$if \: s(t)=s(t+1) \: and \: a(t)=a(t+1)$}\\
       0,& \text{otherwise},
		\end{cases}}
\end{align} 
where $\gamma$ and $\lambda$ are used to discount the sequence of rewards for multiple state and action pairs. Using $T(s(t),a(t),s(t+1),a(t+1))$ and $\mathbb{E}[\Sigma^{\infty}_{t=0} \gamma(t) D_r(t)]$ the agent/s (BS) targets to maximise the expected future return based on the discount factor $(\gamma \in [0,1])$, because $\gamma$ determines the importance between $D_r(t)$ and $D_r(t+1)$ at each timestep$(t_s)$.}

\end{itemize}

To combine the whole concept, in following sections we discuss the reliability from two significant aspects, the first aspect is the tracing method to remember the reliable path as  reliable states $s(t)$ for the future visits at the time slot $(t+1)$. Similarly, the second aspect is based on the DQN part that is assisted by the trace-table $\mathbf{\tau}(s,a)$ to further enhance the reliability. Further details are in following sections:  
\subsubsection{Trace and Allocation}
The optimal policy of the aforementioned parameters can be discovered by an agent using following function:
\begin{align}\label {o}
    \pi^*(s(t))= {\mathop {\arg \max }\limits _{a(t)}Q}(s(t),a(t)),\forall s(t)\in \mathbb {S},
\end{align}
where $\pi^*(s)$ represents the optimal policy. This function provides the optimal policy value for each state $s$ from the finite state set after taking appropriate action $a$.
For a better understanding, the optimal policy can be defined:
\begin{align}
\label {p}  V_{\pi^*}(s(t))&=\myb{{\mathop { \max }\limits _{a(t)}}{\left[r(s(t),a(t))+\gamma\sum_{s(t+1)}{P_{s(t) \rightarrow s(t+1)}V_{\pi^*}(s(t+1))}\right]}},\\
\label{q1-g}\myb{ q_{n}(t)}&\myb{=D{_{r}(t+1)}+\gamma D{_{r}(t+2)}+...+{{\gamma }_{n-1}}D{_{r}(t+n)}+{{\gamma }_{n}}Q({s(t+n)})},
\end{align}
where $n$ denotes the length of the long-term reliability trace, $P_{s(t) \rightarrow s(t+1)}$ shows the transition probabilities from  state $s(t)$ to the next state $s(t+1)$ and the following equitation illustrates the dynamic $\color{black} \lambda(t) =\gamma \mathbf{\tau}(s(t),a(t)) + \nabla V_\pi(s(t),a(t); \mathbf{\theta}) $ based general value policy return (without-$\mathbf{\tau}(s,a)$):
\begin{align}
Q({{s}(t)},{{a}(t)})&\leftarrow Q({{s}(t)},{{a}(t)})+\alpha [{q}_{n}(t)-Q({{s}(t)},{{a}(t)})], \\
{\color{black}{q}^{\lambda }}\color{black}(t)&\color{black}=(1-\lambda(t) )n\sum\limits_{n=1}^{\infty }{{{\lambda}^{n-1}}}{{q}_{n}}(t),
\end{align}
\textcolor{black}{where $(1-\lambda(t))$ is the normalisation factor to ensure that the maximum weights are not more than $1$. Similarly, the term ${\lambda }^{n-1}$ indicates the weighted proportionality of $\lambda$-returns over $n$-steps taken by the agent. The following $a$-values are based on $(n)$-step reliable trace mechanism, in \eqref{q1-t} and \eqref{q2-t}:}
 \myb{\begin{align}
 \label{q2-t} Q(s(t),a(t))&=Q({{s(t)}},{{a(t)}})\leftarrow Q(s(t),a(t))+\alpha \delta (t){{\tau}}(s(t),a(t)),\\
 \label{q1-t}\delta (t)&={{D_r}(t+1)}+\gamma Q({{s}(t+1)},{{a}(t+1)})-Q({{s}(t)},{{a}(t)}),
\end{align}}
The equation \eqref{rt1} shows the initialisation of $\mathbf{\tau}(s,a)$ at initial timestep when $t_s=0$.
\begin{align}\label{rt1}
\mathbb{{\tau}}(s,a)=0,
\end{align}
 \myb{MRP (evolution from $s(t)$ to $s(t+n)$), the overall mapping process for MRP tuple is defines as: 
$\left\{ (s(t),a(t),{D_r}(t)):s(t+1)\sim \mathbb{P}\left( \cdot |s(t),a(t) \right), \\ {{a}(t)}\sim \pi (\cdot |s(t)),{D_r}(t)\sim {D_r}\left( \cdot |s(t),\text{ }a(t) \right) \right\}, \\ {{T}}\in [0,\infty ].$}

\subsubsection{Deep Q Network}
The deep Q network is equipped with the reliability trace table $\mathbf{\tau}(s,a)$, therefore the training network is trained to learn more reliable polices with the help of short experience memory. Similarly, following loss function indicates the accuracy to allocate resources with a reliable policy.   
\begin{align}\label{dqn1}
l(\theta^{DQN}_{train}) = 1/T \sum_{t=1/t \in T}{{{[\omega^{DQN}(t) - Q(s(t),a(t);\theta^{DQN}_{train})] }^{2}}}, 
\end{align}
where
\begin{align}\label{dqn2}
\omega^{DQN}(t) = D_r(t) + \gamma \max \limits _{a(t+1)} ~Q(s(t+1),a(t+1);\theta^{target} (t+1)),
\end{align}
and $\omega^{DQN}_t$ is the target Q-values from target DNN. For the improved training, in general the update frequency of the target network $\theta^{target}$ is performed in slow manner. Due to this reason the target network remains fixed for the target network update threshold $T_e^{'}$.

The DRL agent uses gradient decent method as in (\ref{dqn1}) to reduce the prediction error by minimizing the loss function. The updating of $\theta$ is provided in (16), which is based on the outcome of new experience. The updating function for $\theta$ is defined in (\ref{dqn7}), namely DRL Bellman equation.
\begin{align}
&\label{dqn3}{\theta^{DQN} \gets \theta^{DQN} -[ \omega^{DQN}(t+1)} - Q(s(t),a(t);\theta^{DQN})]\nabla Q(s(t),a(t);\theta^{DQN}),\\
&\label{dqn4}q_{\pi }(s (t), a (t)) = D_r(s (t+1), a (t)) + \gamma \sum \limits _{s (t+1)} \sum \limits _{a (t+1)} {{\myb{P_{s(t) \rightarrow s(t+1)}}}(a)} q_{\pi }(s (t+1), a(t+1)), \\
&\label{dqn5}q_{\pi ^{*}}(s (t),a (t)) = D_r(s (t),a(t)) + \gamma \sum \limits _{s(t+1)} {{\myb{P_{s(t) \rightarrow s(t+1)}}}(a)\max \limits _{a(t+1)} } ~q_{\pi^{*}}(s(t+1),a(t+1)),
\end{align}
where $\nabla Q(s(t),a(t);\theta^{DQN})$ shows the periodic changes in DNN weights and Q-values. Similarly, the function $q_{\pi ^{*}}(s(t),a(t))$ shows Q-values and the long-term reward calculations for the DRL algorithm. That is based on the discount factor $\gamma$ and below mentioned optimal DRL policy $\pi ^{*}$.
\begin{align}\label{dqn6}
{\pi ^{*}}(s(t)) = \mathop {\arg \max }\limits _{a (t)} \left [{ {q_{\pi ^{*}}(s(t),a(t))} }\right], ~\forall ~s\in \mathbb {S},
\end{align}
where $\pi^{*}(s(t))$ represents the optimal policy for the DRL algorithm. This function provides the optimal policy value for each state $s$ from finite sate set after taking appropriate action $a$.
\begin{align}\label{dqn7}
Q(s (t),a (t)) \!\gets \!(1\!-\!\alpha)Q(s(t),a(t)) \!+\!\alpha \!\left [{\! {r(s(t),a(t)) \!+\! \gamma \max \limits _{a (t+1)} Q(s (t+1),a (t+1))} \!}\right]\!,\quad
\end{align}
where $Q(s,a)$ is showing Q-value update according to DRL Bellman equation and the following equation represents the inputs of DNN.
\begin{align}\label{dqn8}
&s(t)=\{a(t)({D_r}(t),{D_r}(t-1)),a(t)({D_r}(t),{D_r}(t-1))\ldots, a(T_t)({D_r}(T_t),{D_r}(T_{t}-1))\}, \\
&\label{weights} \varrho(\vartheta):\sum_{h=0/h \in d_h} W_h \times I_h(s (t)) + B_h,
\end{align}
where $\varrho$ represents activation function with the input as $(\vartheta)$, that is being processed by the hidden layer $h$ with density of neurons $d_h$. Secondly, $W_h$ is weight of the hidden layer $h$ and $I$ shows the inputs plus $B$ is bias term. It is worth noting that the maximum reward budget for each BS is defined as $D_r(\max)\in [0,\inf]$.

\subsection{Algorithmic Details and Descriptions}
This subsection, presents the deep SARSA-$\lambda$ algorithm in \textbf{Algorithm. \ref{alg}} to provide $(n)$-step reliable return for the uplink NOMA-URLLC, while guaranteeing the minimum average system utility as well. In the deep SARSA-$\lambda$ base model, the central control is done by the BS that acts as a deep SARSA-$\lambda$ agent. At each trial the agent (BS) starts observing the current state of a network environment. Which contains the initial 2D association among users, sub-channels for BS denoted by $s(t)$. According to the $s(t)$ and current policy (initial), the agent takes an action $a(t)$ from the action space $\mathbb{A}$. Which represents swapping operation of users from one resource block to an other. After taking the action $a(t)$ by following the greedy policy $\epsilon-greedy$ the agent (BS) receives a feed back reward $D_r(t)$ based on the long-term URLLC objectives \eqref{h}. Therefore, at each trial after observing the state $s$, taking an action $a$ and feed back $D_r(t)$ the Q-value $\mathbf{Q}(s,a)$ and $\mathbf{\tau}(s,a)$ is calculated. In this way, to find the long-term policy the reliability trace and Q-values are updated for each trial. Similarly, each episode represents the long-term optimality of state and action pairs with Q-values and reliability trace $\mathbf{\tau}(s,a)$ as an optimal policy. The main target of the agent is to minimise the long-term error and maximise the data-rate in the form of $D_r(t)$. That is based on the instantaneous average system utility $D_r(t)$.
\par  Finally for the deep SARSA-$\lambda$, according to $\epsilon-greedy$ and the $\mathbf{\tau}(s,a)$ the action $a(t)$ may not be the best choice but for the long-term the optimality of an action is high to achieve long-term reliable allocation policy. In this work, the long-term allocations are considered under continuously changing the URLLC constraints such as packet size$-D$ and user connectivity. Therefore, the exploited scheme is fully online on policy. Base on the $s(t),a(t),D_r(t),s(t+1),a(t+1)$ tuple, the updates of $\mathbf{Q}(s,a)$ are mainly dependent on two significant parts described in the following sections.
\subsubsection{DNN}
\myb{The proposed framework is based on uplink NOMA-URLLC scenarios, therefore the state space is huge $\mathbb{S}$ from which state observations are used as an input to DNN's. The frequent error propagation happens, that results in low reliability. In order to increase the reliability with the minimum complexity, the output of the neural network as $\theta$ weight approximation is performed with following DNN components:}     
\begin{itemize}
\item \underline{\myb{Structure of DNN:}}
\textcolor{black}{We adopted the fully connected DNN structure, the fully connected model is a mapping function of $\mathbb{R}_{layer-n} \rightarrow \mathbb{R}_{layer-m}$. In this way, a fully connected DNN structure learns from all the features of neural network in distributed manner. Therefore, DNN model is applied to explore the wireless network environment efficiently. Lastly, we use sparse activation of neurons to avoid non useful neuron activations \cite{9200330}. }		 
\item \underline{Sparse-Activations:}
We use ReLU activation function fully connected DNN structure to achieve sparsity. Sparsity is one of the main advantage of neuron activations. With this design we minimise the useless neuron activations that helps to reduce the training randomness of neural networks. Secondly, ReLU activations also suffer very less from the log likelihood of the inputs because ReLU saturate in only one way (forward tracing) \cite{glorot2011deep}.
\item \underline{Experience-Memory-$\kappa$:} In conventional DNN's the experience is initialised with randomly taken actions to store in the memory $\kappa$ as an experience. These experiences are based on $s(t),a(t),D_r(t)$ and used as training samples to minimise the training loss between target networks and training networks \eqref{dqn1}.
\item \underline{Optimiser-(Loss):}
The adaptive momentum (ADAM) optimiser is used to approximate the neural network weight training problem by minimising the MSE loss. It is known that ADAM is computationally efficient loss function approximation. Additionally, the proposed $(n)$-step return method produces long-term $(n)$-step gradients with sparsity. The step-size annealing property od ADAM is suitable for the proposed design.
\end{itemize}
In \eqref{weights} a basic neuron activation is represented using activation function $\varrho$ that is based on ReLU activations then for each neuron the sparse output will be as: 
\begin{align}
 \varrho(\vartheta)=\left\{ \begin{matrix}
   \vartheta, & \vartheta\ge 0,  \\
   0, & \vartheta<0,  \\
\end{matrix} \right.
\end{align}

The convergence of ADAM can be expressed with the following regret function:
\begin{align}
  & \kappa (T)=\sum\limits_{t=1}^{T}{[{{\psi }{(t)}}({{\omega_{adam} }{(t)}})-{{\psi }{(t)}}(\tilde{\omega}_{adam} )]},
\end{align}
where $ \tilde{\omega}_{adam} =\arg {{\min }_{\omega \in \chi' }}\sum\limits_{t=1}^{T}{{{\psi }{(t)}}(\omega_{adam} (t))}$ and ${\psi }{(t)}$ is the loss value to calculate the adam weights and gradients.
\subsubsection{Reliability Tracing}
 The proposed framework is initialised with the random network parameters to get initial inputs for deep SARSA-$\lambda$ learning. That includes hyperparameters for learning process, such as $T$, $T_t$, $\gamma$ and $\epsilon-greedy$. After random initial user, BS and sub-channel associations, the reliable deep SARSA-$\lambda$ framework begins to learn the online traces using \eqref{rt1}. We use the $\mathbf{\tau}(s,a)$ mechanism to allocations mechanism and to fill the memory $\kappa$ for the future deep SARSA-$\lambda$ experience replay initialisations instead of random initialisations. Therefore, when the batch is full of reliable experiences from the wireless network environment, then for the further allocations DNN is invoked to predict and allocate. This process is adopted to minimise/avoid the error propagation in DNN's due to the randomness of the initial policy. Therefore, our designed model is based on two networks namely, target network $\theta^*$ and primary network $\theta$. The function of a target network is to train the primary network online by providing the best allocation policy that assists the primary network to learn and compare the correctness of current and future allocations. Similarly, to validate the current allocation policy error minimisation is performed to get the final long-term outcomes.  
The memory $\kappa$ updates are based on reliable actions $\mathbf{\tau}(s,a)$ then the loss for $(n)$-step experience-based DNN's can be expressed as follows:
\begin{multline}
l(\theta^{DQN}_{train}) = (s(t),a(t),D_r(t),s(t+1),a(t+1),\mathbf{\tau}(s,a))\sim (\kappa)\\ \lbrace{1/T \sum_{t=1/t \in T}{{{[\omega^{DQN}(t)- Q(s(t),a(t);\theta^{DQN}_{train})] }^{2}}}\rbrace}. 
\end{multline}

   
\begin{algorithm}[tp!]
\footnotesize
\caption{Deep SARSA$-\lambda$ with Reliability Tracing Mechanism}
\label{alg}
\begin{algorithmic}[1]
\State Inputs:
\begin{enumerate}
\item $T$ Max-Episodes
\item $T_t$ Max-Trial
\item $\alpha$ Step-size
\item $\gamma$ Discount-factor
\item $\epsilon-greedy$ Policy
\end{enumerate}
\State Initialization parameters:
\begin{enumerate}
\item Communication system ($M,D,\mathbf{C},\mathbf{P},N_s,s_j,N_u,u_k$)
\item Q-Table $\mathbf{Q}(s,a)$
\item Trace-Table $\mathbf{\tau}(s,a)$
\item Memory $\kappa$
\item $\lambda=0.99$ 
\end{enumerate}	
    \State $\mathbb{S}$, $\mathbb{A}$
    \State Random user association to $BS$ and any $c$
%
\For {$t_s$ = $1$:$T_{}$}
\State  $s_1=rand()$
\State Random power allocations $P$
  \For {j = $t$:$T_t$}
       \State $s(t),a(t)$
       \State  Update $\myb{D_r(t)=
       \begin{cases}
       \sum\limits_{k=1}^{N_{u}^{j}}R_{k}^{j}(t),& \text{$if \ \mathbb{E}\left[\sum\limits_{j=1}^{N_s}\sum\limits_{k=1}^{N_{u}^{j}}\,{\varepsilon_{k}^j} (t-1)\right]\ge \mathbb{E}\left[\sum\limits_{j=1}^{N_s}\sum\limits_{k=1}^{N_{u}^{j}}\,{\varepsilon_{k}^j} (t)\right]$} \\& \text{ $and$ \ $\sum\limits_{j=1}^{N_s}\sum\limits_{k=1}^{N_{u}^{j}}{c_k^j(t-1)=\sum\limits_{j=1}^{N_s}\sum\limits_{k=1}^{N_{u}^{j}}{c_k^j(t)}}$}\\
       0,& $otherwise.$
		\end{cases}}$ 
       \State compute \myb{$q_{\pi ^{*}}(s(t),a(t)) = D_r(s(t),a(t))+\gamma \sum \limits _{s(t+1)} {P_{s(t) \rightarrow s(t+1)}(a)\max \limits _{a(t+1)} } ~q_{\pi ^{*}}(s(t+1),a(t+1)).$}
       \State compute $\omega^{DQN}(t) = D_r(t) + \gamma \max \limits _{a(t+1)} ~Q(s(t+1),a(t+1);\theta^{target} (t+1))$, and $l(\theta^{DQN})= 1/T \sum_{t=1}^{T}{{{[\omega^{DQN}(t) -Q(s(t),a(t);\theta^{DQN})] }^{2}}}, $
       \State  $(s(t),s(t+1)), (a(t),a(t+1))$ 
  \If {$T_t > State-size$ }
 	\State fetch $(s(t),s(t+1)), (a(t),a(t+1))$ from previous Experience $\mathbf{\tau}(s,a)$ 
 \EndIf
  \EndFor   	
\EndFor
    \For {i = $1$:$\kappa$}
    
        \For {j = $1$:$\kappa$}
        	\State $s(t),a(t)$
        	\State Update $      \myb{D_r(t)=
       \begin{cases}
       \sum\limits_{k=1}^{N_{u}^{j}}R_{k}^{j}(t),& \text{$if \ \mathbb{E}\left[\sum\limits_{j=1}^{N_s}\sum\limits_{k=1}^{N_{u}^{j}}\,{\varepsilon_{k}^j} (t-1)\right]\ge \mathbb{E}\left[\sum\limits_{j=1}^{N_s}\sum\limits_{k=1}^{N_{u}^{j}}\,{\varepsilon_{k}^j} (t)\right]$} \\& \text{ $and$ \ $\sum\limits_{j=1}^{N_s}\sum\limits_{k=1}^{N_{u}^{j}}{c_k^j(t-1)=\sum\limits_{j=1}^{N_s}\sum\limits_{k=1}^{N_{u}^{j}}{c_k^j(t)}}$}\\
       0,& $otherwise.$
		\end{cases}}$              	
        	       				          		       				        						
        	\State Calculate $Q({s(t)},{a(t)})\leftarrow Q({s(t)},{a(t)})+\alpha [{q}_{n}(t)-Q({s(t)},{a(t)})]$
        	\State update $\color{black}\lambda(t) =\gamma \mathbf{\tau}(s(t),a(t)) + \nabla V_\pi(s(t),a(t); \mathbf{\theta}) $
        	\State Calculate $\myb{\mathbf{\tau}(s,a)=\begin{cases}
       \gamma\lambda\tau(s(t-1),a(t-1))+1, &   \text{$if \: s(t)=s(t+1) \: and \: a(t)=a(t+1)$}\\
       0,& \text{otherwise},
		\end{cases}}$
        	\State Search $\pi^*(s)$
        	\State $T(s(t),a(t),s(t+1),a(t+1))/p(s',D_r|s,a)=\Pr \{{s(t)}={s'},{D_R(t)}=D_r(t)|{s(t-1)}=s,{a(t-1)}=a\}$
        	\State {$\mathbf{\tau}(s,a) \leftarrow \gamma \lambda \tau(s,a)$}
        	\State  $(s(t),s(t+1)), (a(t),a(t+1))$       	
		\EndFor		
   	        \State $\mathbf{\tau}(s,a) \rightarrow DQN(\theta)$
   	        \State Add new Experience$\rightarrow \kappa$
   \EndFor
	\State $\mathbf{\tau}(s,a)$
\State Return Reliable Weighted Allocation $\theta$ 	
\end{algorithmic}
\end{algorithm}

\subsection{Analysing Proposed Framework}
\subsubsection{Convergence analysis}
Prior to start discussion on convergence analysis of the proposed deep SARSA-$\lambda$, it is necessary to discuss/know the convergence details for traditional reinforcement learning algorithms. Subsequently, following by the discussion on convergence of traditional reinforcement learning algorithms the convergence for the proposed model can be easily verified.
The traditional reinforcement learning, that is $(1)$step learning algorithm converges by satisfying $\alpha \in \lbrace{0,1\rbrace}$ to the wel-known Q-function based on bellman equations. The detailed prove is already given in \cite{melo2001convergence}. 
\par The deep SARSA$-\lambda$ algorithm uses $\lambda$ and $\mathbf{\tau}(s,a)$ to learn the best path for the future allocation policy that results in faster learning. Similarly, combining the additional power of DNN proposed model is capable to converge even faster to the optimal state and action pair.  
\subsubsection{Complexity analysis}
\myb{The complexity of proposed framework is based on the number of user communicating and the total number of clusters/resource blocks. The experiments in our propose framework are performed with two different examples, for the first case $\lbrace{N_s=5,N_u=5\rbrace}$ and for the second case $\lbrace{N_s=5,N_u=7\rbrace}$. The computation complexity for our framework is based on the trace-table $\mathcal{O}(\mathbf{\tau}(\mathbb {S} \times \mathbb {A}))$, $\mathbb{S}=\lbrace{N_u^{N_s}\rbrace}$ total number of states, and $\mathbb {A}=(N_{s}-1) \times N_s$ represents total number of actions for each state. Therefore, the total complexity is as (trace-table + Weighted matrix of DQN), as $\mathcal{O}(\mathbf{\tau}(\mathbb {S} \times \mathbb {A})+\mathbf{Q}(s,a;\theta^{DQN}))$. Where the complexity of trace table is based on the number of states and action that is calculated as $\mathcal{O}(\mathbf{\tau}(\mathbb {S} \times \mathbb {A}))$. The weighted matrix is experience memory that is based on the experiences from \eqref{dqn8}. Therefore the worst case complexity of the weighted matrix is computed as $\mathcal{O}((\theta))$. Furthermore, the complexity of the bias and weighted terms are calculated separately as $\mathcal{O}(|\mathbb {S}|d_h(|\mathbf{B_h}|+|\mathbf{W_h}|))$, with $d_h$ total number of neurons in each layer.}

\begin{table}[htb!]
\small
\centering
\caption{Network settings}
\label{tab1}
\myb{\begin{tabular}{|l||l|}
\hline
   Max Trials &  500   \\ \hline
   Max Time steps &  500   \\ \hline
   Total BSs &  1   \\ \hline
   Total resource blocks &  5   \\ \hline
   Experience Memory &  500   \\ \hline
   Batch size &  500   \\ \hline
   Optimiser &  Q-table, ADAM   \\ \hline
   Hidden layers &  2  \\ \hline
   Number of neurons in each layers &  500 \\ \hline
   Neuron activation &  ReLu   \\ \hline
   Type of DNN structure &  Fully-Connected (Fig.\ref{frame}) \\ \hline
   DNN-performance &  Loss-function   \\ \hline
   Dynamic-Reward &  Data-rate (t)  \\ \hline
   Loss-calculations &  MSE (mean squared error)   \\ \hline
   Optimiser-batch &  500   \\ \hline
   Max user for each resource block &  $(5)$, $(7)$   \\ \hline
   Bandwidth &  $1$ MHz  \\ \hline
   $\alpha$ &  4  \\ \hline
   Traffic Types based on $D$ &  $8$ \\ \hline
   $\sigma^2$ &  $-174$   \\ \hline
   $\gamma_{\delta}$  &$0.6$  \\ \hline
   Learning rate $\alpha_r$ & $0.75$   \\ \hline
   $\epsilon-greedy$ & $0.01$   \\ \hline
   $(n)$-step $\lambda$ & Dynamic \\ 
\hline
\end{tabular}}
\end{table}
\section{Simulation Results}
The main aim of this section is to show the validity of a proposed framework that is based on the Deep SARSA$-\lambda$. Similarly, the convergence of the proposed Deep SARSA$-\lambda$ algorithm illustrates the validity of the performance enhancement. An efficient convergence of the Deep SARSA$-\lambda$ agent proves the long-term reliable resource allocation assistance in uplink NOMA-URLLC networks. Furthermore, to illustrate the effectiveness we test the performance of the proposed algorithm with 4 communication scenarios. The first scenario is based on NOMA-URLLC network with bursty network traffic, in which the packet size for each user is variable. The second scenario is a fixed packet NOMA-URLLC network that consists of a fixed packet size for each user. Similarly, the third and fourth scenarios are based on different network loads, such as the different number of users connected to each resource block. Therefore, for the third scenario, we allow $(2-5)$ and in the fourth scenario, the network load is $(2-7)$ for each resource block. Additionally, in simulations at the initial timestep, the random allocation is adopted to start the allocation enhancement framework. Lastly, TABLE \ref{tab1} contains the details for the simulation parameters. 

Our simulation setup was based on Intel core i7-7700 CPU with 3.60 GHz frequency having 16 GB of RAM (Random Access Memory) and 64-bit operating system (windows-10). All the experiments are simulated using Python 3.6. From TABLE II for both the algorithms we used 500 timesteps with 500 timeslots for each episodes.

\begin{figure}[t!]
  \centering
\includegraphics[scale=.8,keepaspectratio]{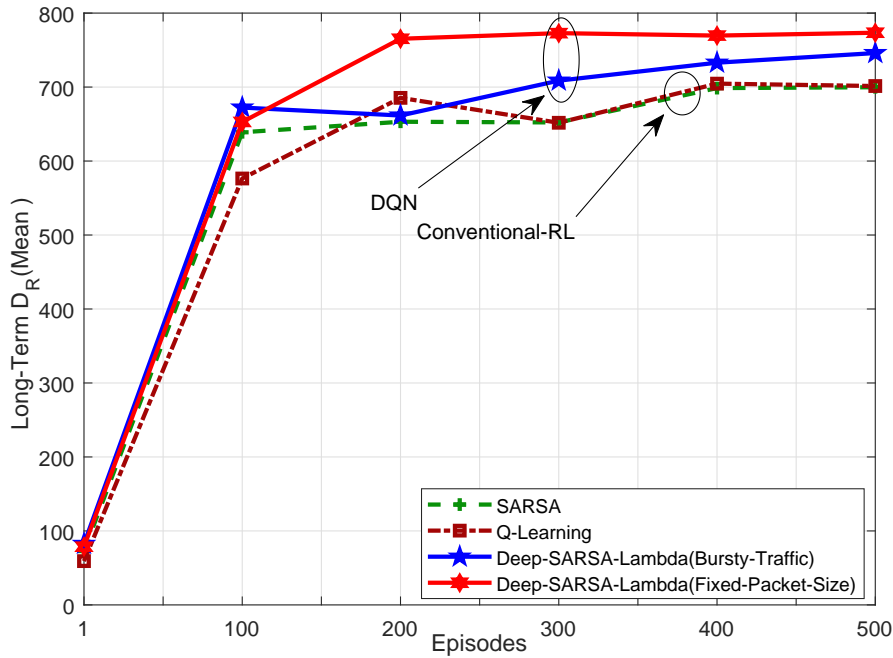}
  \caption{ \myb{Illustrating the comparison between two convergence parameters, long-term mean decoding error and long-term reward (expectation-sum-rate) over the number of episodes. Presenting the convergence of deep SARSA-$\lambda$ framework with variable network traffic (bursty) in blue coloured plot, static (fixed packet size) network traffic in red coloured plot, traditional Q-learning in dotted maroon coloured line and SARSA Q-learning with green line. Where the network is based on $1$ BS, $N_{s}=5$ sub-channels, and each sub-channel can serve maximum of $N_{s}=5$ network users at most.}}\label{1}
\end{figure}
    
\subsection{\myb{Convergence Rate $(D_r)$}}
\myb{Fig. \ref{1} characterizes the inter-correlations between achieved rewards and the number of episodes. In this figure, we compare the proposed algorithm with conventional Q-learning and SARSA Q-learning. It is apparent from the figure that the convergence of the traditional Q-learning and SARSA Q-learning algorithm is poor. This is due to the extremely huge state and action space for the time varying wireless network model. Therefore, the traditional RL agents are presenting unreliable random outcomes for the bursty network traffic. Due to the power of DNN, $\mathbf{\tau}(s,a)$, and dynamic $\lambda$, the proposed deep SARSA-$\lambda$ algorithm is able to converge with better rewards in $200$ episodes for NOMA network environments. In the start, the agent is trying to explore the environment by taking random actions. Due to this, the performance is worst in this case. Similarly, in $100$ episodes the agent has gained significant experience to exploit the environment. Therefore, in $100$ episodes the reward increases rapidly and after $200$ episodes the reward is more stable. Which represents the stable performance of the proposed agent with better rewards after gaining further experiences from the environment. Additionally, this figure also shows the inverse effect of the bursty network traffic on convergence. On the other hand, convergence is much faster and better for static network traffic with fixed packet size. This is because a dynamically changing network increases the overall allocation complexity. Lastly, the proposed reward function consists of real-time reward feedback that is represented by the average rate for all NOMA-URLLC users. Moreover, the scale of the state and action space is based on the maximum connections allowed for each resource blocks and total number of available resource blocks. Therefore, it is denoted by $\mathbb{S}=\lbrace{N_u^{N_s}\rbrace}$ and $\mathbb {A}=(N_{s}-1) \times N_s$ represents total number of actions for each state.}    

\begin{figure}[t!]
  \centering
  \includegraphics[scale=.8,keepaspectratio]{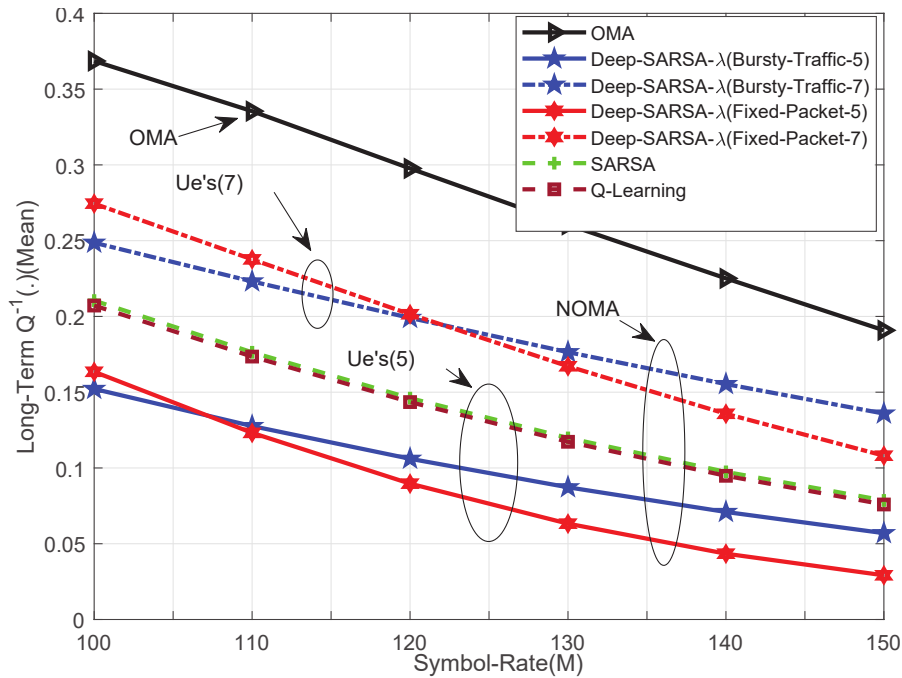}
  \caption{\myb{Showing the comparison among decoding error (mean), symbol rate $(M)$, type of network traffic, network density, NOMA and conventional OMA for deep SARSA-$\lambda$, Q-learning and SARSA Q-learning.}}\label{2}
\end{figure}

\subsection{\myb{Long-Term Error Performance With respect to The Symbol Rate }}
\myb{In Fig. \ref{2} obtained outcomes summarise some of the main characteristics of the long-term relationship among network error probability, symbol rate (M), network traffic density, type of traffic, OMA, NOMA, benchmark schemes, and the proposed algorithm. The most striking result to emerge from the figure is that, for the network density of $(5)$ users, the average error is less than $7$ users. Similarly, for the bursty type of traffic when the symbol rate is smaller than $110$ the error rate is efficient for $5$ users. For $7$ users, the error rate is higher for the bursty traffic for all the cases except $M\leq130$. Therefore, the results obtained from the preliminary analysis of these parameters indicate that the mean error of the FBL transmission is directly proportional to the network density and inversely proportional to the symbol rate $(M)$. It is apparent from this figure that for all settings when the symbol rate $(M)$, is increasing then the average error is decreasing and vice versa. Interestingly, the performance for the long-term mean error probability of the deep SARSA-$\lambda$ algorithm is more efficient than traditional Q-learning methods. Similarly, as compared to conventional OMA for all network settings the proposed learning-based NOMA system is performing efficiently.}
\begin{figure}[t!]
  \centering
\includegraphics[scale=.8,keepaspectratio]{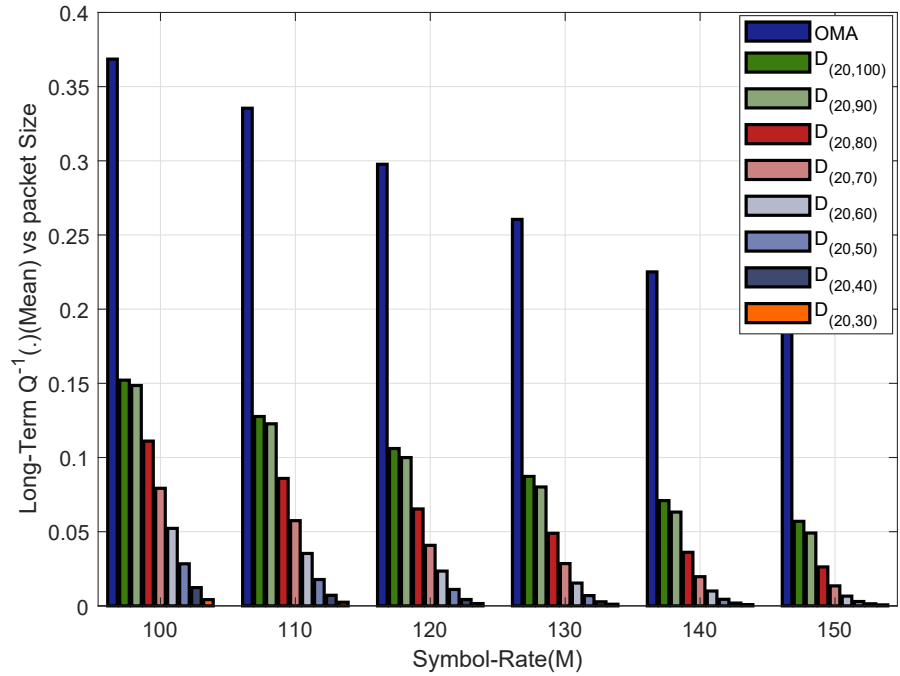}
  \caption{Presenting the comparison among, packet size $D$, symbol rate (M), decoding error (mean) with NOMA and OMA techniques.}\label{3}
\end{figure}

\subsection{\myb{Long-Term Error Rate For The Packet Size and Symbol Rate}}
In Fig. \ref{3} we present the breakdown of long-term error, variable packet size $(D)$, and symbol rate $(M)$. This figure is quite revealing in several ways. Firstly, the average error rate for the NOMA-URLLC is much better than the conventional OMA-URLLC scheme. Secondly, the average error rate is decreasing when the symbol rate is increasing. Similarly, when the symbol rate is decreasing, then the average error starts increasing. Lastly, the packet size $D$ is also inversely proportional to the long-term error. Due to this, when the packet size $(D)$ increases from $(20,30)$ to $(20,100)$ bits then the error rate starts increasing. 
      
\begin{figure}[t!]
  \centering
  \subfigure[]{\includegraphics[scale=.8,keepaspectratio]{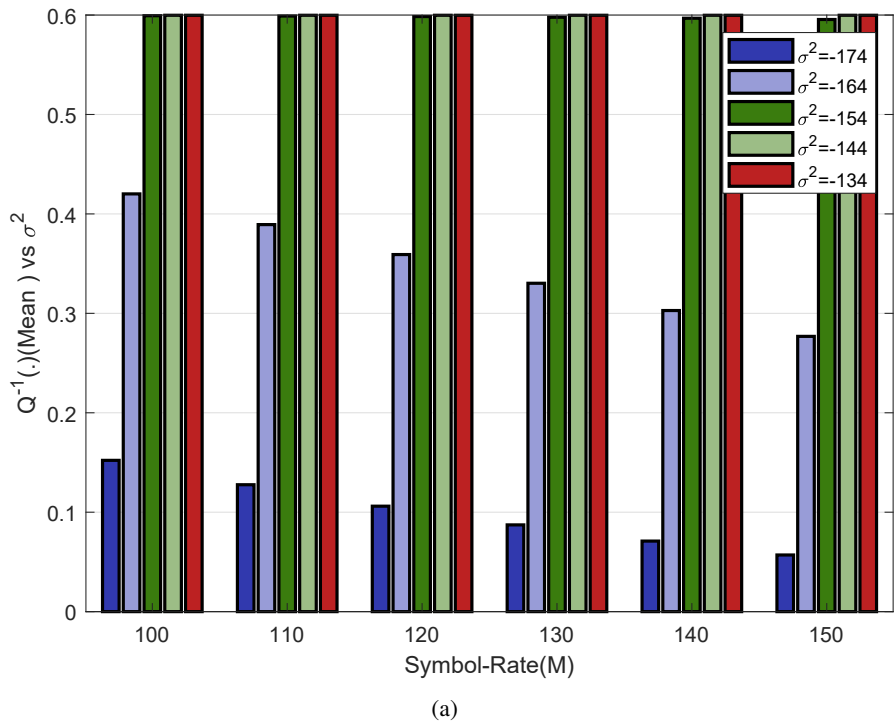}}
  \caption{Illustrates the comparison of decoding error (mean) with $\sigma^2$ and symbol rate (M).}\label{4}
\end{figure}

\subsection{\myb{Long-Term Error Performance With $\sigma^2$ and Symbol rate }}
In Fig. \ref{4} we plot the summary statistics for the comparison among long-term error,$\sigma^2$, and Symbol rate. As the figure portrays, there is a significant difference between the two groups, when $(\sigma^2=-174)$ and $(\sigma^2 \geq -164 )$. The average error is small for the case when $\sigma^2=-174$ but it increases more than $50 \%$ for $\sigma^2 \geq -164$ case. Therefore, if the noise is stronger then $-164$, it results in high error rates. Therefore, the results from this figure indicate that regardless of the increased symbol rate $(M)$ the blockage starts when $\sigma^2 \geq -164 $ as a result, the average error increases significantly.  
\begin{figure}[t!]
  \centering
\includegraphics[scale=.8,keepaspectratio]{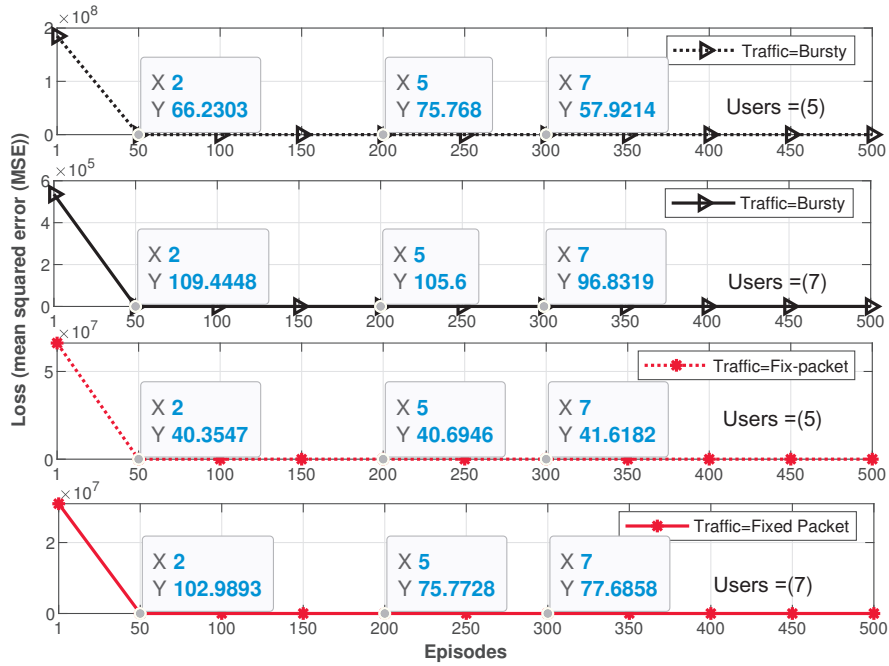}
  \caption{\myb{Shows the mean squared error (MSE) as a DNN loss function for two different network densities having (5) and (7) maximum users for each sub-channel with two different types of network traffic.}}\label{6}
\end{figure}

\subsection{\myb{Loss Function}}
The results in Fig. \ref{6} compares the preliminary analysis of DNN loss function (MSE), episodes, network density and type of traffic. From this figure, we can see that the prediction loss for the static network traffic with low density is less as compared to the other cases. Therefore, by converging within $100$ episodes, the loss stays below $42$ during the whole simulation time. Similarly, the DNN prediction loss starts increasing when the type of traffic switches to burst traffic, which is due to the increased network complexity. Which also becomes higher for the increased network density. Therefore, users at each resource block are increased from $(5)$ \myb{users} to $(7)$ \myb{users}. The DNN prediction loss is increasing. However, due to the power of DNN reliable trace $\mathbf{\tau}(s,a)$ and sparse activations, the proposed DNN structure is capable to handle network dynamics at the cost of negligible performance. The proposed framework provides an efficient long-term allocation policy for different types of networks with $(109)$ maximum average loss.  
\begin{figure}[t!]
  \centering
\includegraphics[scale=.8,keepaspectratio]{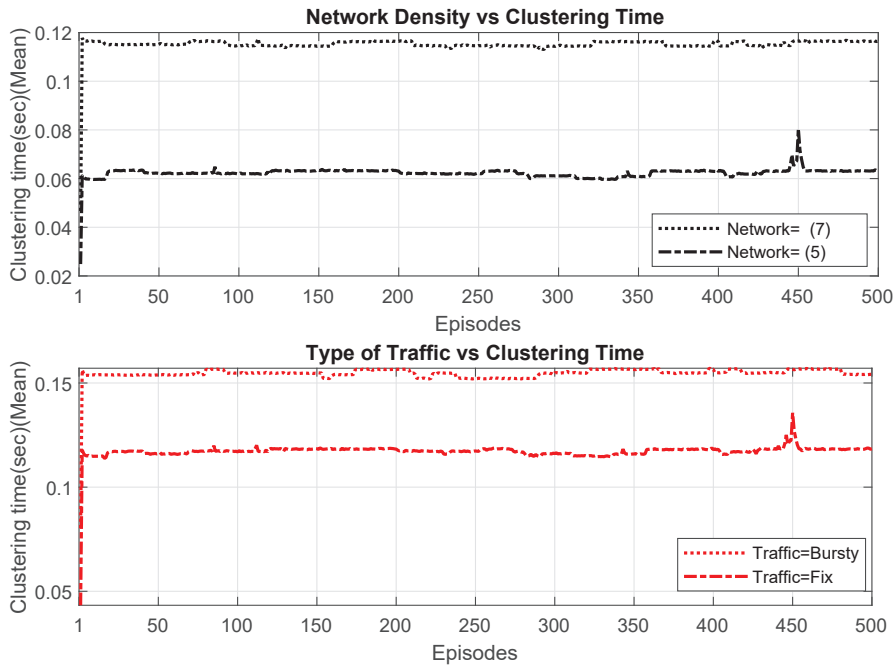}
  \caption{Shows the mean clustering time (sec) for two different network loads and two different types of network traffic.}\label{71}
\end{figure}

\subsection{\myb{Clustering Time}} 
In Fig. \ref{71} we plot the comparison of the clustering time, network density and type of traffic. Similarly, the top sub-figure compares the clustering time between network density and episodes. In the top sub-figure, the clustering time starts increasing when the network load for each resource block is increased. Hence, there is proportional behaviour between network density and clustering time. The bottom sub-figure illustrates the clustering time comparison between episodes and type of traffic. It is quite revealing that the bursty network traffic scenario costs more time due to complex resource allocations. However, static network traffic scenarios are showing opposite behaviour with less clustering time. Therefore, the type of traffic is inversely proportional to the DNN loss (MSE) and clustering time, as shown in Fig. \ref{6}. Lastly, when we compare both of the sub-figures, another interesting trend depicts that the clustering time increases simultaneously by changing the type of traffic. However, the clustering time is low for increased network loads. Therefore, we can say that change in network traffic costs more clustering time and allocation complexity as compared to the variable network loads.         
\section{Conclusion}
In this paper, we proposed a long-term mean error minimization framework for uplink NOMA-URLLC networks. For all considered URLLC settings, NOMA is a more efficient scheme than traditional OMA. In particular, NOMA achieved $70\%$ long-term mean error performance gain over the OMA case. With the aid of the reward function, the proposed DRL approach outperforms traditional models within \myb{$200$} episodes. To prove the efficiency of our solution, we compared the long-term mean error performance of the proposed deep SARSA-$\lambda$ with two well-known reinforcement learning algorithms namely, Q-learning and SARSA Q-learning. The results indicated that the proposed approach achieves $37\%$ long-term mean error performance gain over the traditional Q-learning and $38\%$ better than the SARSA Q-learning algorithm. 
\bibliographystyle{IEEEtran}
\bibliography{single2}

\begin{thebibliography}{10}
\providecommand{\url}[1]{#1}
\csname url@samestyle\endcsname
\providecommand{\newblock}{\relax}
\providecommand{\bibinfo}[2]{#2}
\providecommand{\BIBentrySTDinterwordspacing}{\spaceskip=0pt\relax}
\providecommand{\BIBentryALTinterwordstretchfactor}{4}
\providecommand{\BIBentryALTinterwordspacing}{\spaceskip=\fontdimen2\font plus
\BIBentryALTinterwordstretchfactor\fontdimen3\font minus
  \fontdimen4\font\relax}
\providecommand{\BIBforeignlanguage}[2]{{%
\expandafter\ifx\csname l@#1\endcsname\relax
\typeout{** WARNING: IEEEtran.bst: No hyphenation pattern has been}%
\typeout{** loaded for the language `#1'. Using the pattern for}%
\typeout{** the default language instead.}%
\else
\language=\csname l@#1\endcsname
\fi
#2}}
\providecommand{\BIBdecl}{\relax}
\BIBdecl

\bibitem{123456789}
W.~Ahsan, W.~Yi, Y.~Liu, and A.~Nallanathan, ``Reliable reinforcement learning
  based {NOMA} schemes for {URLLC},'' in \emph{{IEEE} Glob. Commun. Conf.
  {(GLOBECOM)}}, Dec. 2021.

\bibitem{elbayoumi2020noma}
M.~Elbayoumi, M.~Kamel, W.~Hamouda, and A.~Youssef, ``{NOMA}-assisted
  machine-type communications in{UDN}: State-of-the-art and challenges,''
  \emph{{IEEE} Commun. Surveys Tuts.}, vol.~22, no.~2, pp. 1276--1304, 2020.

\bibitem{xiang2020noma}
Z.~Xiang, W.~Yang, Y.~Cai, Z.~Ding, Y.~Song, and Y.~Zou, ``{NOMA}-assisted
  secure short-packet communications in {IoT},'' \emph{{IEEE} Wireless
  Commun.}, vol.~27, no.~4, pp. 8--15, 2020.

\bibitem{zhang2020sparse}
X.~{Zhang}, L.~{Yang}, Z.~{Ding}, J.~{Song}, Y.~{Zhai}, and D.~{Zhang},
  ``Sparse vector coding-based multi-carrier{ NOMA} for in-home health
  networks,'' \emph{{IEEE} J. Sel. Areas Commun.}, vol.~39, no.~2, pp.
  325--337, 2021.

\bibitem{qian2020noma}
L.~Qian, Y.~Wu, F.~Jiang, N.~Yu, W.~Lu, and B.~Lin, ``{NOMA} assisted
  multi-task multi-access mobile edge computing via deep reinforcement learning
  for industrial internet of things,'' \emph{{IEEE} Trans. Ind. Informat.}, pp.
  1--1, 2020.

\bibitem{mu2020exploiting}
W.~{Yi}, Y.~{Liu}, A.~{Nallanathan}, and M.~{Elkashlan}, ``Clustered
  millimeter-wave networks with non-orthogonal multiple access,'' \emph{IEEE
  Trans. Commun.}, vol.~67, no.~6, pp. 4350--4364, Jun. 2019.

\bibitem{gui20206g}
G.~Gui, M.~Liu, F.~Tang, N.~Kato, and F.~Adachi, ``{6G}: Opening new horizons
  for integration of comfort, security and intelligence,'' \emph{{IEEE}
  Wireless Commun.}, vol.~27, no.~5, pp. 126--132, 2020.

\bibitem{jiang2019packet}
X.~Jiang, Z.~Pang, M.~Zhan, D.~Dzung, M.~Luvisotto, and C.~Fischione, ``Packet
  detection by a single {OFDM} symbol in {URLLC} for critical industrial
  control: A realistic study,'' \emph{{IEEE} J. Sel. Areas Commun.}, vol.~37,
  no.~4, pp. 933--946, 2019.

\bibitem{khoshnevisan20195g}
M.~Khoshnevisan, V.~Joseph, P.~Gupta, F.~Meshkati, R.~Prakash, and
  P.~Tinnakornsrisuphap, ``{5G} industrial networks with{ CoMP for URLLC and}
  time sensitive network architecture,'' \emph{{IEEE} J. Sel. Areas Commun.},
  vol.~37, no.~4, pp. 947--959, 2019.

\bibitem{xiang2019concurrent}
L.~Xiang, R.~G. Maunder, and L.~Hanzo, ``Concurrent{ OFDM }demodulation and
  turbo decoding for ultra reliable low latency communication,'' \emph{{IEEE}
  Trans. Veh. Technol.}, vol.~69, no.~2, pp. 1281--1290, 2019.

\bibitem{she2017radio}
C.~She, C.~Yang, and T.~Q. Quek, ``Radio resource management for ultra-reliable
  and low-latency communications,'' \emph{{IEEE} Commun. Mag.}, vol.~55, no.~6,
  pp. 72--78, 2017.

\bibitem{polyanskiy2010channel}
Y.~Polyanskiy, H.~V. Poor, and S.~Verd{\'u}, ``Channel coding rate in the
  finite blocklength regime,'' \emph{{IEEE} Trans. Inf. Theory}, vol.~56,
  no.~5, pp. 2307--2359, 2010.

\bibitem{yang2014quasi}
W.~Yang, G.~Durisi, T.~Koch, and Y.~Polyanskiy, ``Quasi-static multiple-antenna
  fading channels at finite blocklength,'' \emph{{IEEE} Trans. Inf. Theory},
  vol.~60, no.~7, pp. 4232--4265, 2014.

\bibitem{shahraki2021comprehensive}
A.~Shahraki, M.~Abbasi, M.~Piran, M.~Chen, S.~Cui \emph{et~al.}, ``A
  comprehensive survey on {6G} networks: Applications, core services, enabling
  technologies, and future challenges,'' \emph{arXiv preprint
  arXiv:2101.12475}, 2021.

\bibitem{xiao2019downlink}
C.~Xiao, J.~Zeng, W.~Ni, X.~Su, R.~P. Liu, T.~Lv, and J.~Wang, ``Downlink
  {MIMO-NOMA }for ultra-reliable low-latency communications,'' \emph{{IEEE} J.
  Sel. Areas Commun.}, vol.~37, no.~4, pp. 780--794, 2019.

\bibitem{amjad2018performance}
M.~Amjad and L.~Musavian, ``Performance analysis of{ NOMA}for ultra-reliable
  and low-latency communications,'' in \emph{{Proc. IEEE Glob. Commun. Conf.
  (GLOBECOM)}}, 2018, pp. 1--5.

\bibitem{kotaba2019improving}
R.~Kotaba, C.~N. Manchon, N.~M.~K. Pratas, T.~Balercia, and P.~Popovski,
  ``Improving spectral efficiency in {URLLC via NOMA-Based Retransmissions},''
  in \emph{Proc. IEEE Int. Commun. Conf. (ICC)}, 2019, pp. 1--7.

\bibitem{chen2019optimal}
J.~Chen, L.~Zhang, Y.-C. Liang, and S.~Ma, ``Optimal resource allocation for
  multicarrier {NOMA} in short packet communications,'' \emph{{IEEE} Trans.
  Veh. Technol.}, vol.~69, no.~2, pp. 2141--2156, 2019.

\bibitem{8933345}
H.~{Ren}, C.~{Pan}, Y.~{Deng}, M.~{Elkashlan}, and A.~{Nallanathan}, ``Joint
  power and blocklength optimization for {URLLC} in a factory automation
  scenario,'' \emph{{IEEE} Trans. Wireless Commun.}, vol.~19, no.~3, pp.
  1786--1801, 2020.

\bibitem{maraqa2019survey}
O.~Maraqa, A.~S. Rajasekaran, S.~Al-Ahmadi, H.~Yanikomeroglu, and S.~M. Sait,
  ``{A survey of rate-optimal power domain NOMA schemes for enabling
  technologies of future wireless networks},'' \emph{arXiv preprint
  arXiv:1909.08011}, 2019.

\bibitem{huang2020deep}
H.~Huang, Y.~Yang, Z.~Ding, H.~Wang, H.~Sari, and F.~Adachi, ``Deep
  learning-based sum data rate and energy efficiency optimization for
  {MIMO-NOMA }systems,'' \emph{{IEEE} Trans. Wireless Commun.}, vol.~19, no.~8,
  pp. 5373--5388, 2020.

\bibitem{ye2019deep}
N.~Ye, X.~Li, H.~Yu, A.~Wang, W.~Liu, and X.~Hou, ``{Deep learning aided
  grant-free NOMA toward reliable low-latency access in tactile Internet of
  Things},'' \emph{{IEEE} Trans. Ind. Informat.}, vol.~15, no.~5, pp.
  2995--3005, 2019.

\bibitem{9310298}
L.~Huang, L.~Zhang, S.~Yang, L.~P. Qian, and Y.~Wu, ``Meta-learning based
  dynamic computation task offloading for mobile edge computing networks,''
  \emph{{IEEE} Commun. Lett.}, vol.~25, no.~5, pp. 1568--1572, 2021.

\bibitem{ye2020deepnoma}
N.~Ye, X.~Li, H.~Yu, L.~Zhao, W.~Liu, and X.~Hou, ``{DeepNOMA: A unified
  framework for NOMA using deep multi-task learning},'' \emph{{IEEE} Trans.
  Wireless Commun.}, vol.~19, no.~4, pp. 2208--2225, 2020.

\bibitem{fu2019dynamic}
Y.~Fu, W.~Wen, Z.~Zhao, T.~Q. Quek, S.~Jin, and F.-C. Zheng, ``{Dynamic power
  control for NOMA transmissions in wireless caching networks},'' \emph{{IEEE}
  Wireless Commun. Lett.}, vol.~8, no.~5, pp. 1485--1488, 2019.

\bibitem{parvez2018survey}
I.~Parvez, A.~Rahmati, I.~Guvenc, A.~I. Sarwat, and H.~Dai, ``A survey on low
  latency towards {5G: RAN}, core network and caching solutions,'' \emph{{IEEE}
  Commun. Surveys Tuts.}, vol.~20, no.~4, pp. 3098--3130, 2018.

\bibitem{azari2019risk}
A.~Azari, M.~Ozger, and C.~Cavdar, ``{Risk-aware resource allocation for
  URLLC}: Challenges and strategies with machine learning,'' \emph{{IEEE}
  Commun. Mag.}, vol.~57, no.~3, pp. 42--48, 2019.

\bibitem{she2020tutorial}
C.~She, C.~Sun, Z.~Gu, Y.~Li, C.~Yang, H.~V. Poor, and B.~Vucetic, ``{A
  tutorial of ultra-reliable and low-latency communications in {6G}:
  Integrating theoretical knowledge into deep learning},'' \emph{arXiv preprint
  arXiv:2009.06010}, 2020.

\bibitem{hernandez2019understanding}
J.~F. Hernandez-Garcia and R.~S. Sutton, ``{Understanding multi-step deep
  reinforcement learning: A systematic study of the DQN target},'' \emph{arXiv
  preprint arXiv:1901.07510}, 2019.

\bibitem{silver2016mastering}
D.~Silver, A.~Huang, C.~J. Maddison, A.~Guez, L.~Sifre, G.~Van Den~Driessche,
  J.~Schrittwieser, I.~Antonoglou, V.~Panneershelvam, M.~Lanctot \emph{et~al.},
  ``Mastering the game of go with deep neural networks and tree search,''
  \emph{nature}, vol. 529, no. 7587, pp. 484--489, 2016.

\bibitem{mnih2015human}
V.~Mnih, K.~Kavukcuoglu, D.~Silver, A.~A. Rusu, J.~Veness, M.~G. Bellemare,
  A.~Graves, M.~Riedmiller, A.~K. Fidjeland, G.~Ostrovski \emph{et~al.},
  ``Human-level control through deep reinforcement learning,'' \emph{nature},
  vol. 518, no. 7540, pp. 529--533, 2015.

\bibitem{watkins1992q}
C.~J. Watkins and P.~Dayan, ``{Q}-learning,'' \emph{Machine learning}, vol.~8,
  no. 3-4, pp. 279--292, 1992.

\bibitem{sutton2018reinforcement}
R.~S. Sutton and A.~G. Barto, \emph{Reinforcement learning: An
  introduction}.\hskip 1em plus 0.5em minus 0.4em\relax MIT press, 2018.

\bibitem{ahsan2021resource}
W.~Ahsan, W.~Yi, Z.~Qin, Y.~Liu, and A.~Nallanathan, ``{Resource Allocation in
  Uplink NOMA-IoT Networks: A Reinforcement-Learning Approach},'' \emph{{IEEE}
  Trans. Wireless Commun.}, 2021.

\bibitem{shao2020significant}
Y.~Shao, A.~Rezaee, S.~C. Liew, and V.~W. Chan, ``Significant sampling for
  shortest path routing: A deep reinforcement learning solution,'' \emph{{IEEE}
  J. Sel. Areas Commun.}, vol.~38, no.~10, pp. 2234--2248, 2020.

\bibitem{kiani2018edge}
A.~Kiani and N.~Ansari, ``Edge computing aware {NOMA} for {5G} networks,''
  \emph{{IEEE} Internet Things J.}, vol.~5, no.~2, pp. 1299--1306, 2018.

\bibitem{8680645}
F.~{Fang}, Z.~{Ding}, W.~{Liang}, and H.~{Zhang}, ``Optimal energy efficient
  power allocation with user fairness for uplink {MC-NOMA} systems,''
  \emph{{IEEE} Wireless Commun. Lett.}, pp. 1--1, 2019.

\bibitem{8807386}
J.~{Cui}, Y.~{Liu}, and A.~{Nallanathan}, ``Multi-agent reinforcement
  learning-based resource allocation for {UAV} networks,'' \emph{{IEEE} Trans.
  Wireless Commun.}, vol.~19, no.~2, pp. 729--743, 2020.

\bibitem{9200330}
D.~C. Nguyen, P.~Cheng, M.~Ding, D.~Lopez-Perez, P.~N. Pathirana, J.~Li,
  A.~Seneviratne, Y.~Li, and H.~V. Poor, ``{Enabling AI in Future Wireless
  Networks: A Data Life Cycle} perspective,'' \emph{{IEEE} Commun. Surveys
  Tuts.}, vol.~23, no.~1, pp. 553--595, Firstquarter 2021.

\bibitem{glorot2011deep}
X.~Glorot, A.~Bordes, and Y.~Bengio, ``Deep sparse rectifier neural networks,''
  in \emph{AISTATS}, 2011, pp. 315--323.

\bibitem{melo2001convergence}
F.~S. Melo, ``Convergence of {Q}-learning: A simple proof,'' \emph{Institute Of
  Systems and Robotics, Tech. Rep}, pp. 1--4, 2001.

\end{thebibliography}
\end{document}